\documentclass[12pt]{article}
\usepackage{graphicx} 
\graphicspath{ {./pics/} }
\usepackage{hhline}
\usepackage{amssymb}
\usepackage{amsmath}
\usepackage{booktabs}
\usepackage{siunitx}
\usepackage{times}
\usepackage{cite}
\usepackage{lineno}
\usepackage{url}
\usepackage{chngcntr}
\usepackage{csquotes}
\usepackage[english]{babel}
\usepackage{indentfirst}
\usepackage{float}
\usepackage{placeins}
\usepackage{xcolor}
\usepackage[x11names]{xcolor}
\usepackage{amsmath}
\usepackage{bbding}

\usepackage[dvipsnames]{xcolor}

\usepackage[a4paper, total={6in, 9in}]{geometry}

\usepackage{hyperref}
\hypersetup{
    colorlinks=true,
    linkcolor=magenta,
    urlcolor=blue,
    citecolor=blue,
    pdftitle={Overleaf Example},
    }
\begin{document}


\vspace*{2cm}

\begin{center}
\begin{Large}

{\boldmath \bf
    Production of ${\Lambda}$ hyperons in 4.0A~GeV and 
                4.5A~GeV carbon-nucleus interactions at the Nuclotron
}

\vspace{0.5cm}

BM$@$N Collaboration

\end{Large}
\end{center}

\vspace{0.5cm}

{\noindent
S.\,Afanasiev, G.\,Agakishiev, A.\,Aleksandrov, E.\,Aleksandrov, I.\,Aleksandrov, P.\,Alekseev, K.\,Alishina, V.\,Astakhov, T.\,Aushev, V.\,Azorskiy
% ...
}

\newpage

\begin{abstract}
\noindent

\par The BM@N experiment (Baryonic Matter at the Nuclotron) is the first fixed-target 
experiment at the JINR NICA accelerator complex. 
In this work, data on the interactions of a carbon-ion beam with kinetic energies of  
4.0A~GeV and 4.5A~GeV with C, Al, Cu, and Pb targets are used to measure transverse 
momentum spectra and rapidity distributions of $\Lambda$ hyperon yields.
The results are compared with the predictions of DCM-SMM, UrQMD, and PHSD transport 
models and with the $\Lambda$ yield measurements in other experiments at similar 
collision energies.

\end{abstract}

\newpage


\section{Introduction}
\label{sect1}

The study of collisions of relativistic nuclei provides a unique opportunity to 
explore nuclear matter under extreme conditions of high density and temperature. 
Theoretical models suggest that beam kinetic energies in the range of $5-20$A~GeV 
(corresponding to $\sqrt{s_{NN}}=3.6 - 6.4$~GeV) are optimal for exploring the 
hadronization phase transition and the properties of dense baryonic matter ~\cite{Arsene}.

The Nuclotron at the NICA accelerator complex provides a wide range of ion beams 
within the energy range of $\sqrt{s_{NN}}=2.3 - 3.5$~GeV. These energies are sufficiently 
high for the production of strange mesons and (multi-) strange hyperons in nucleus--nucleus 
collisions close to the kinematical threshold. Measurements in this regime are therefore 
essential for understanding the mechanism of strangeness production and the dynamics of 
the dense hadronic matter ~\cite{NICAWhitePaper,BMN_CDR}.

\(\Lambda\) hyperons, which contain a single strange quark, are important observables 
in this context. Their production near threshold is sensitive to multi-step processes, 
secondary interactions, and in-medium effects. Their transverse momentum spectrum and 
rapidity distribution provide insight into the degree of thermalization, the system 
dynamics and the underlying reaction mechanisms in heavy-ion collisions. 

The production of $\Lambda$ hyperons at low and intermediate energies has been studied 
by several experiments, such as
FOPI (GSI)~\cite{FOPI_K0_Lambda}, HADES (GSI)~\cite{HADES2010}, and 
STAR (RHIC)~\cite{STAR_Strangeness}. These experiments provided detailed information 
on $\Lambda$ hyperon production in symmetric or almost symmetric systems such 
as Ni~+~Ni, Ar~+~KCl, and Au~+~Au. At higher energies, the collider experiments 
STAR (RHIC) and ALICE (LHC) have investigated hyperon production primarily around 
midrapidity ~\cite{STARPolarization, ALICEPolarization, STAR_BESPolarization, STAR_BESCentral}. 
However, systematic measurements of $\Lambda$ hyperon production in light-ion collisions 
on light and heavy nuclear targets at few-GeV energies in both the central and forward 
rapidity regions remain limited. In particular, asymmetric systems at near-threshold energies 
have not been extensively explored. 

The BM@N experiment is the first fixed-target experiment operating at the NICA (JINR) 
accelerator complex, designed to investigate dense baryonic matter in heavy-ion collisions 
at Nuclotron energies. Thanks to its fixed-target configuration, the BM@N enables studies 
of both symmetric and asymmetric beam-target combinations over a broad rapidity  range, 
including forward kinematics. At the moment the BM@N has collected data on collisions 
of carbon, argon, krypton, and xenon beams with different solid targets.

Results on the production of $\pi^+$ and $K^+$ mesons ~\cite{The_BM@N_collab_kaons_pions} 
and of protons, deuterons, and tritons ~\cite{The_BM@N_collab_protons_deuterons_tritons} 
in argon-nucleus interactions were recently published. 

This paper presents results on $\Lambda$ hyperon  production  in carbon-nucleus 
interactions at the beam kinetic energies of 4.0A~GeV and 4.5A~GeV ($\sqrt{s_{NN}}=3.32$ 
and ${3.46}$~GeV). Transverse momentum and rapidity spectra are measured within BM@N 
acceptance, and $\Lambda$ yields and production cross-sections are extracted and 
extrapolated to the full phase space. The results are compared with predictions of 
the DCM-SMM, UrQMD, and PHSD transport models, as well as with measurements from other 
experiments at similar collision energies.  

\par The paper is organized as follows: Section~\ref{sect2} outlines the BM@N 
experimental setup. Section~\ref{sect3} details the Monte Carlo simulations 
and trigger efficiency evaluation.
Sections ~\ref{sect4}~--~\ref{sect6} cover the 
analysis methodology and the event selection procedure. 
Section~\ref{sect7} describes the procedure 
for the evaluation of the cross-sections, yields and estimation of the 
systematic uncertainties. The results and their comparison with models predictions 
and other experiments are summarized in Section~\ref{sect8}.


\section{Experimental setup}
\label{sect2}

\begin{figure}[tbh]

\begin{center}
\vspace{-0.2cm}

\includegraphics[width=13cm, height=6cm] {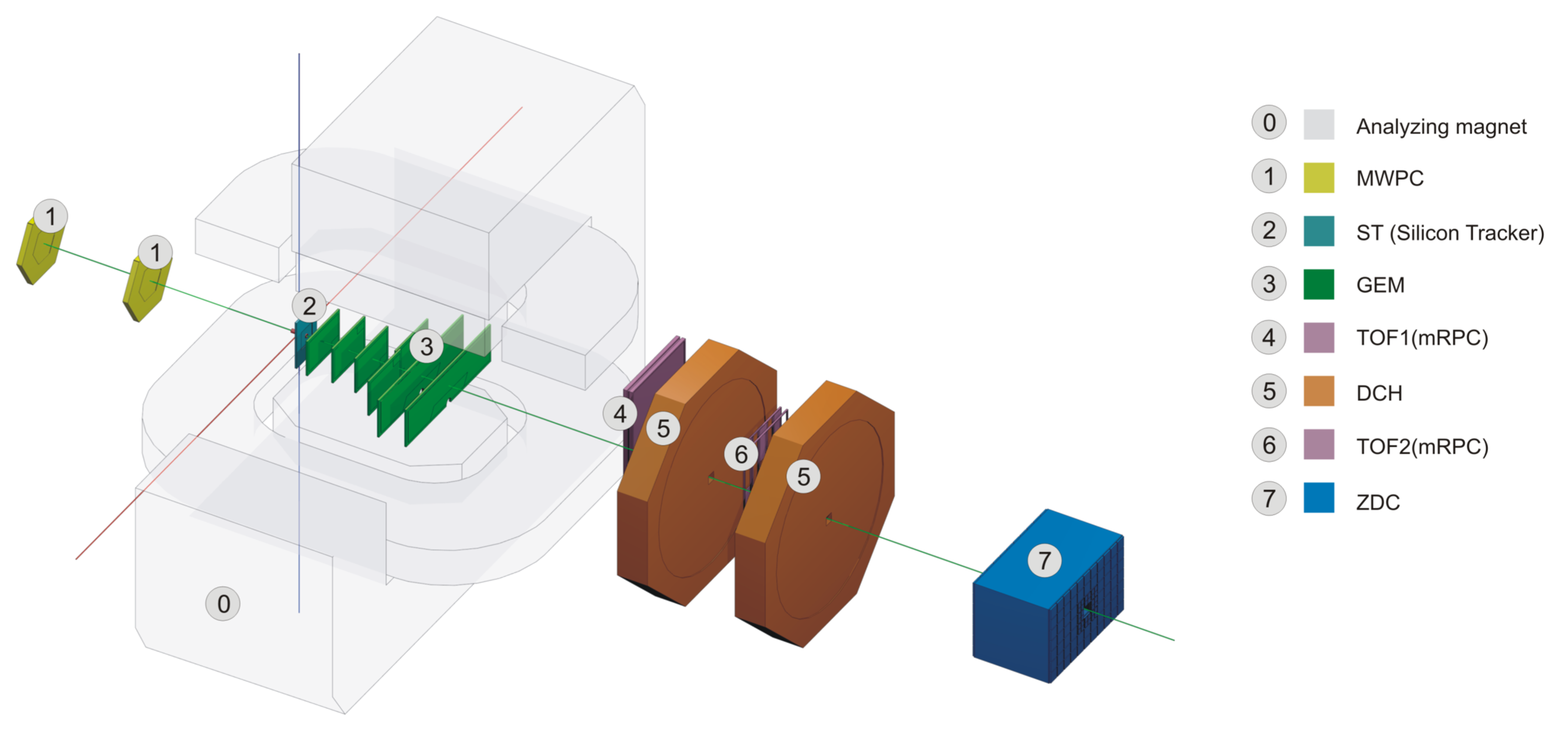}

\end{center}

\caption{Scheme of the BM@N setup in the carbon beam run.}
\label{fig:01}
\end{figure}

The data used in this analysis were collected with a carbon beam with kinetic energies 
of 4.0A~GeV and 4.5A~GeV colliding with four different targets: C, Al, Cu, and Pb. 
The BM@N setup in the carbon-nucleus run is shown schematically in Figure~\ref{fig:01}. 

The BM@N spectrometer is a forward magnetic spectrometer, consisting of beam detectors, 
a central tracker system located inside a dipole analyzing magnet ($0.61T$) and outer 
detectors (not used in this analysis). Charged-particle tracking in this analysis was 
performed using the central tracking system consisting of one plane of a forward silicon 
tracking detector (ST) with double-sided readout and six two-coordinate 
GEM (Gaseous Electron Multiplier) stations composed of five GEM detectors of
 $66\times41~\mathrm{cm}^2$ size and two GEM detectors of $163\times45~\mathrm{cm}^2$ size \cite{BMN_GEM}. 

The tracking stations were arranged such that the beam passed through their centers.
Each successive GEM station was rotated by $180^o$ around the vertical axis. This was 
done to ensure opposite electron drift direction in the successive stations in order
to avoid a systematic shift of reconstructed tracks due to the Lorentz 
angle in the magnetic field. 
A technical description of the BM@N spectrometer is provided 
in ~\cite{The BM@N spectrometer, BMNproject}.

In the present analysis, information from the forward Si detector and GEM 
detectors ~\cite{BMN_Cher,BMN_QM} was used to reconstruct tracks, as well as primary 
and secondary vertices. Since the GEM tracker configuration was tuned to measure 
relatively high-momentum particles, the geometric acceptance for the relatively soft
 decay products of strange $V0$ ($\Lambda$, $K_S^0$) particles was rather low (a few percent). 
 The acceptance covers mostly the forward rapidity region in the laboratory frame. 
 The limited acceptance is accounted for in the analysis through detailed Monte Carlo 
 simulations and bin-by-bin acceptance corrections, as described in Section~\ref{sect5}.

\begin{figure}[tbh]

\begin{center}
\vspace{-0.8cm}

\includegraphics[width=13cm, height=8cm]{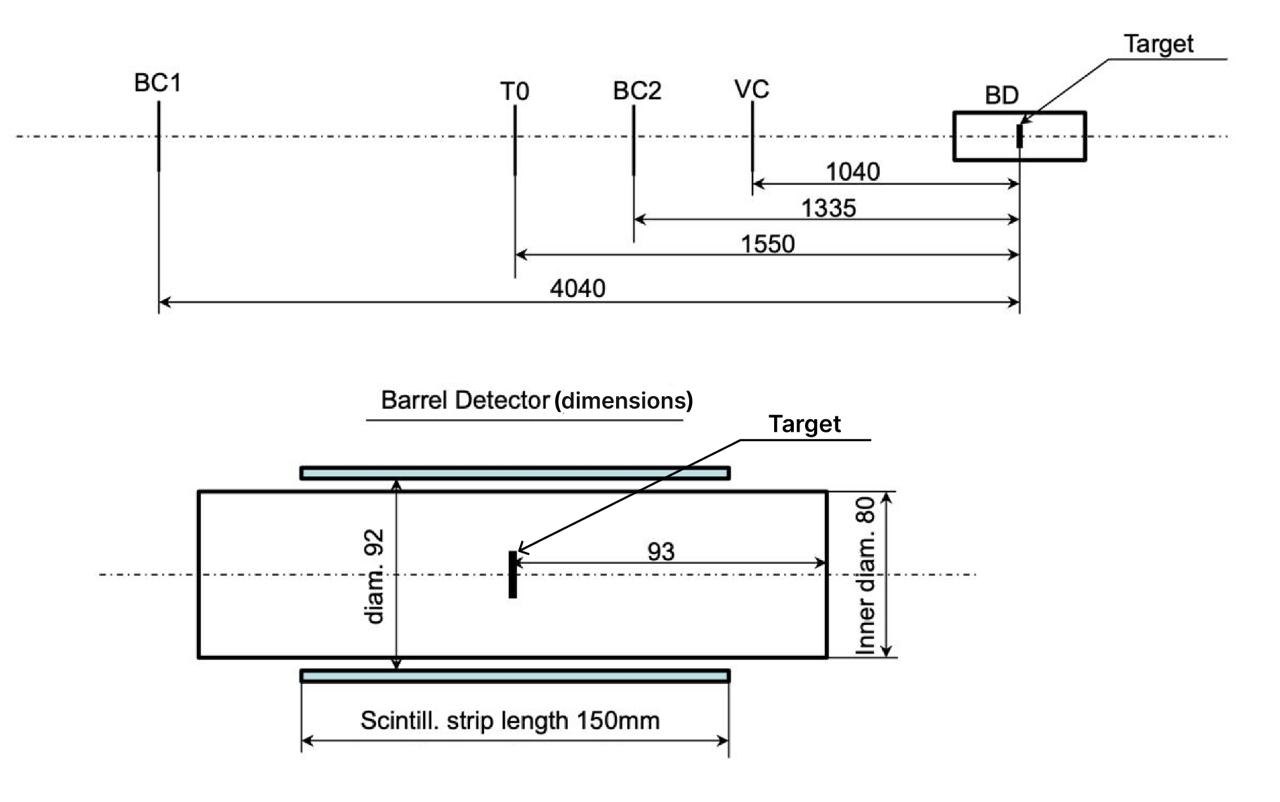}

\end{center}
\vspace{-1.0cm}

\caption{Schematic view of the beam counters, barrel detector, and target position. 
The target was installed in the center of the cylindrical barrel detector, on its axis.  
All dimensions are given in millimeters.}

\label{fig:02}
\end{figure}

\indent The minimum-bias trigger was based on the charged-particle multiplicity measured 
in the cylindrical Barrel Detector (BD), surrounding the target as shown in Figure~\ref{fig:02}. 
The trigger condition required the hit multiplicity in the BD detector exceeds the threshold 
value, which was adjusted for the C, Al, Cu, and Pb targets. Additional beam-monitoring 
and timing information were provided by the beam counters (BC2), veto counter (VC), 
and T0 system (Figure~\ref{fig:02}).
The carbon beam intensity during data taking was of the order of a few $10^5$ 
particles per spill, with a spill duration of $2.0-2.5$ seconds. After quality 
selection, the analyzed data sample comprises approximately 13 million events at
4.0A~GeV and 16 million events at 4.5A~GeV.

\section{Monte Carlo simulations}
\label{sect3}

Monte Carlo event samples of C~+~A collisions were produced with the 
DCM-SMM event generator~\cite{DCM_QGSM,DCM_SMM}. The particle transport through 
the setup volume was simulated with the GEANT4 library~\cite{GEANT4} integrated 
into the BmnRoot software framework~\cite{BmnRoot}. 

The GEM detector response in the magnetic field was simulated with the
micro-simulation package Garfield++~\cite{Garfield}.
The package provides a detailed description of the processes inside the GEM detector,
including the drift and diffusion of released electrons in the electric and magnetic
fields, as well as electron multiplication in the GEM foils. As a result, the output signal from 
the readout plane is well reproduced \cite{BMN_GEM}. To speed up the simulation, the dependencies
of the Lorentz shifts and the charge distributions on the readout planes were 
parameterized and used in the GEM digitization part of the BmnRoot package. 
The details of the detector alignment and Lorentz shift corrections are described 
in ~\cite{BMN_PPNL}. 

The trigger efficiency $\varepsilon_{\mathrm{trig}}$ was evaluated with the DCM-SMM  model 
simulated data. To reproduce the BD trigger response, the BD multiplicity distributions 
were constructed by combining simulated events containing $\Lambda$ hyperons 
with $\delta$-electron background events, which were identified as the main source of 
fake BD signals in C~+~A interactions. The BD trigger condition was applied to the 
Monte Carlo reconstructed events within the BM@N tracking system acceptance. 
The trigger efficiency is defined as a number of $\Lambda$ events with passing 
BD criteria ${N^{\mathrm{rec}}_{\Lambda}(\mathrm{BD} \ge n)}$ to the total reconstructed 
number of $\Lambda$ events ${N^{\mathrm{rec}}_{\Lambda}}$:

\begin{equation}
\varepsilon_{\mathrm{trig}} =
\frac{N^{\mathrm{rec}}_{\Lambda}(\mathrm{BD} \ge n)}
{N^{\mathrm{rec}}_{\Lambda}}.
\end{equation}

The calculated trigger efficiency ranges from $(80 \pm 2)\%$ for C~+~C to $(95 \pm 2)\%$ 
for minimum-bias C~+~Pb interactions.

The systematic uncertainties of $\varepsilon_{\mathrm{trig}}$ include:
\begin{enumerate}
    \item The contribution of the $\delta$-electron background evaluated by 
          varying the effective target thickness in the simulations between $0.5$ 
          and $1.0$ of the nominal target thickness;
    \item The spread of the trigger efficiency values due to variations of 
          the rapidity $y$ and transverse momentum $p_T$ intervals of reconstructed $\Lambda$ hyperons;
    \item Variations of the trigger efficiency caused by reweighting the 
          simulated track multiplicity distributions to match the experimental ones.
\end{enumerate}

The trigger efficiency obtained in the simulation was cross-checked using data samples recorded with reduced trigger requirements. No significant dependence of the trigger efficiency on the event multiplicity was observed within the studied range, indicating that the centrality bias of the trigger is negligible.


\section{Event selection}
\label{sect4}

\par Charged particle track reconstruction was based on the cellular automaton 
approach ~\cite{Kisel}. Events with  $\Lambda$ hyperon decay were selected by the 
candidate-driven approach, in which pairs of oppositely charged tracks form reconstructed secondary vertices.

$\Lambda$ hyperons were identified via their dominant weak decay channel, 
$\Lambda \rightarrow p+\pi^-$. Particle identification was not used in this analysis. 
Therefore, all positive tracks were considered as protons, while negative ones were 
considered as $\pi^-$. Contamination from other particle species contributes to the 
combinatorial background and is accounted for by the invariant-mass background-subtraction 
procedure described below.

The track selection criteria were as follows:
\begin{itemize}

\item Each track was required to have hits in at least four of six GEM stations, where 
      a hit is defined as a combination of two strip clusters on both readout sides 
      ($X$ and $X'$ views) of each detector~\cite{BMN_GEM}.

\item The momentum range of the positive tracks is limited to 
      $p_{pos}< 3.9 (4.4)$~GeV/c at 4.0 (4.5)A~GeV carbon beam energy to suppress 
      contributions from beam fragments.

\item The momentum range of negative tracks is limited to
      $p_{neg}> 0.3$~GeV/c.

\item The distance of closest approach ($DCA$) between two tracks in 
      the $X-Y$ plane at the vertex reconstructed $Z$ coordinate is smaller than $1.0$~cm.

\item The distance between the primary and secondary vertices (decay path length~-~$path$) 
      is required to be larger than $2.5$~cm.

\end{itemize}

\begin{figure}[H]
    \centering
    \includegraphics[width=0.65\textwidth]{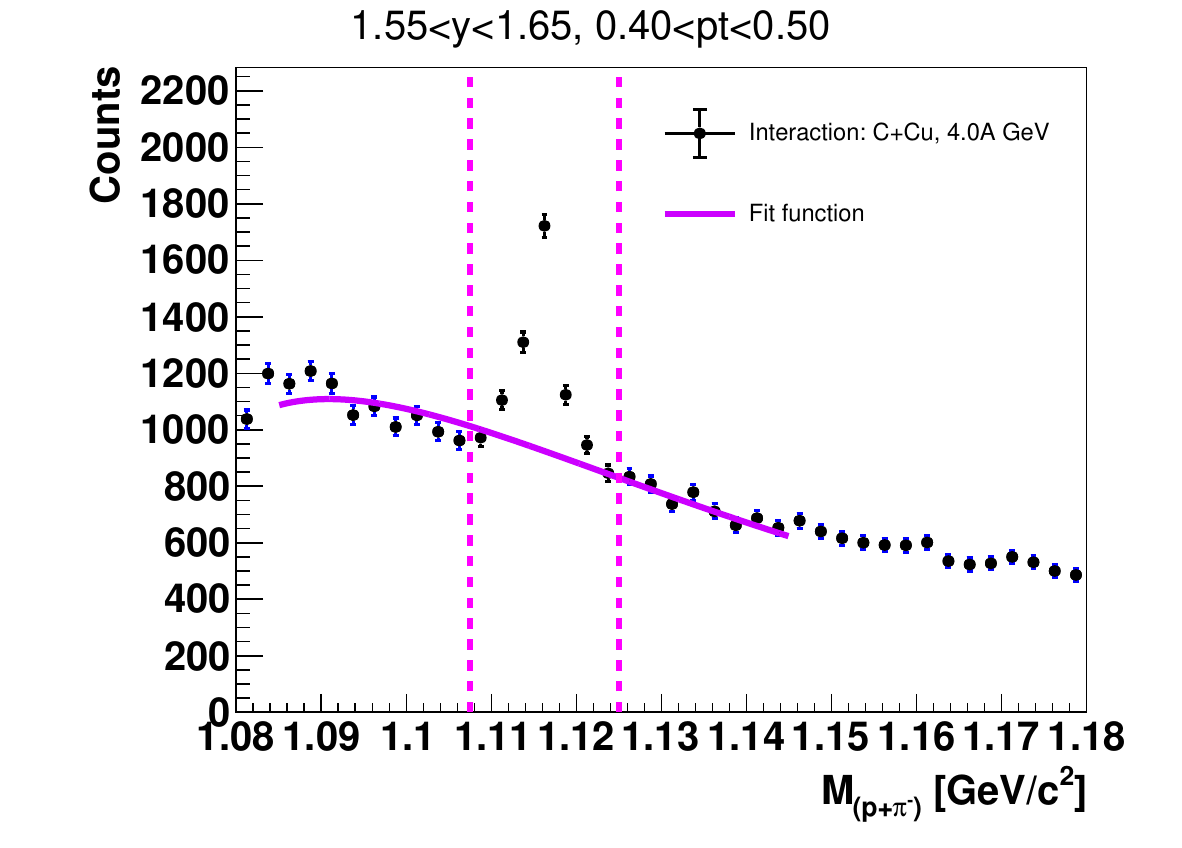}
    \caption{Invariant mass spectrum of $(p,\pi^-)$ pairs reconstructed from Monte 
    Carlo-generated events in the 4.0A~GeV carbon beam with the Cu target (black points). 
    The distribution corresponds to the kinematic region with rapidity $1.55<y<1.65$ 
    and transverse momentum $0.4<p_{T}<0.5~GeV/c$. The purple solid line represents 
    the result of the background fit according to Eq.~(2) (see text for details).
    The magenta dashed vertical lines denote the mass window where the $\Lambda$ 
    signal is calculated as an excess of events with respect to the background fit.}
    \label{lambreco}
\end{figure}

The signal from $\Lambda$ hyperon decays is observed as a narrow peak in the invariant mass 
distribution of two oppositely charged tracks  with the proton and pion mass hypothesis, 
as shown in Figure~\ref{lambreco} for a representative $(y,p_{T})$ interval.

\FloatBarrier

The invariant mass distribution in each kinematic interval was modeled as the sum of a 
signal component and a combinatorial background contribution. The signal peak was 
described by a Gaussian function, $P(m,M_s,\sigma) \propto exp(-0.5\cdot ((m-M_s) / \sigma)^2)$, 
where $m$ is the running invariant mass, $M_s$ and $\sigma$ are free fit parameters and 
denote the peak position and width, respectively. The background was parameterized by a 
threshold function multiplied by an exponential term:

\begin{equation}
f_{\text{bg}}(m) = N \cdot (m - M_0)^A \cdot \exp\left( -B \cdot (m - M_0) \right),
\end{equation}

\noindent where $N$, $A$, and $B$ are free parameters of the fit function  
and $M_0$ = 1.078 GeV is the kinematic threshold. 

The peak region within $\pm  2.5\sigma$ around the $\Lambda$ mass $M_s$ was excluded 
from the background fit to avoid signal distortion. The $\Lambda$ signal was then 
defined as the sum of the bin contents within the peak region after the background 
subtraction. The same signal-extraction procedure was consistently applied to 
Monte Carlo and experimental data.


\section{Acceptance evaluation}
\label{sect5}

The geometrical acceptance and reconstruction efficiency (referred to as detector 
acceptance thereafter) for $\Lambda$ hyperons were evaluated using the DCM-SMM model 
data generated for the C beam at 4.0A~GeV and 4.5A~GeV energies and C, Al, Cu, and Pb 
targets. The simulated data were processed with the same reconstruction and analysis 
chain as the experimental data.

The kinematic range considered in the analysis ($1.2<y<2.1$ and $0.1<p_T<1.05$~GeV/$c$) was divided 
into 8~×~8 cells in the $(y,~p_{T})$ space~\cite{Alishina}. In each cell, the signal was evaluated 
following the procedure described in Section~\ref{sect4}. 

The detector acceptance value $\omega_{acc}$ was calculated as the ratio of the number 
of reconstructed $\Lambda$  hyperons to the number of generated $\Lambda$ 
in each ($y,~p_T$) cell:~$\omega_{acc}$= N$_{rec_{MC}}$/N$_{gen_{MC}}$. 

\begin{figure}[tbh] 
\begin{center}
\hspace{-0.5cm}
\includegraphics[width=7.0cm, height=5.4cm]{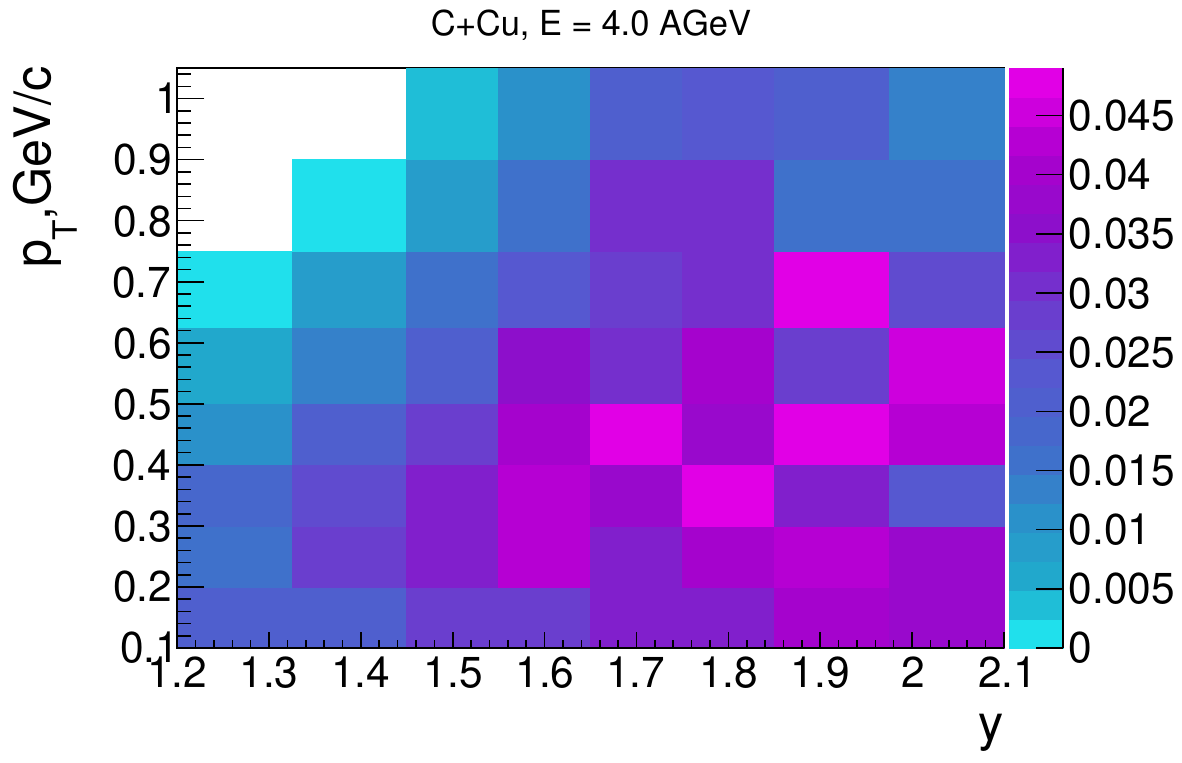}
\includegraphics[width=7.0cm, height=5.4cm]{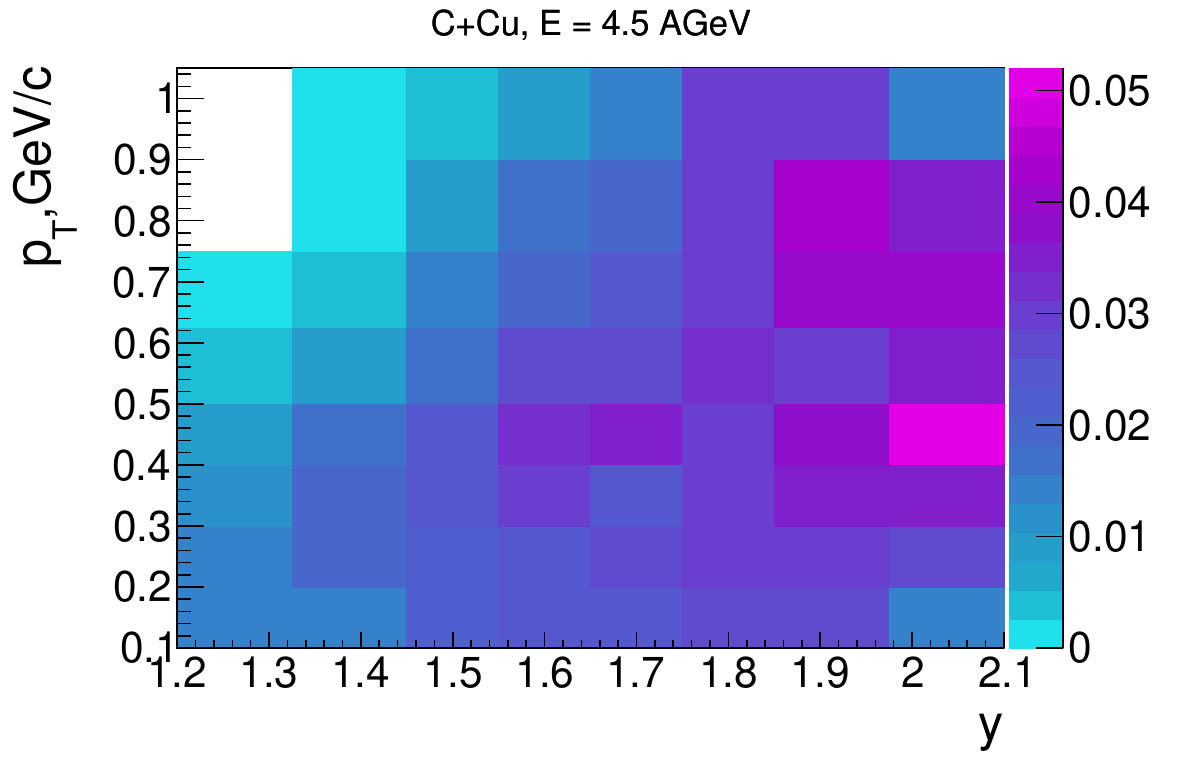}
\end{center}
\vspace{-0.8cm}
 \caption{Detector acceptance for $\Lambda$ emitted in given rapidity and transverse 
          momentum intervals calculated for C~+~Cu interactions at the beam energy of 
	  4.0A~GeV (left) and 4.5A~GeV (right).
 }
 \label{lamb_reco}
\end{figure}

The resulting acceptance maps for C~+~Cu interactions at 4.0A~GeV and 4.5A~GeV are shown in
 Figure~\ref{lamb_reco} as representative examples. The white areas indicate cells with negligible
  efficiencies. All cells with acceptance below $0.01$ were excluded from the analysis.
An extrapolation procedure was performed to obtain the number of expected $\Lambda$ events for these low acceptance regions according to the model.

The extrapolation factor $f_{extrap}$ at a given $p_T$ was calculated as the ratio of the 
number of MC-generated $\Lambda$ hyperons in all the cells along the ${p_{T}}$ direction 
for a given rapidity interval to the number of reconstructed $\Lambda$ hyperons with 
reconstruction efficiency $\omega_{acc_i} > 0.01$ in that rapidity interval.

The systematic uncertainties of the calculated acceptance were evaluated using a bootstrap 
sampling method. For each $(y,~p_{T})$ cell, the invariant mass distribution was resampled 
1000 times; each time, the signal value was calculated. The distribution of reconstructed 
yields was parameterized with a Gaussian function, and the $\sigma$ of the fit was used 
as the systematic uncertainty of N$_{rec_{MC}}$ evaluation. The full acceptance uncertainty 
includes both statistical and systematic contributions.


\section{Data analysis}
\label{sect6}

\begin{figure}[tbh]
\begin{center}
\vspace{-0.2cm}
\includegraphics[width=0.45\textwidth]{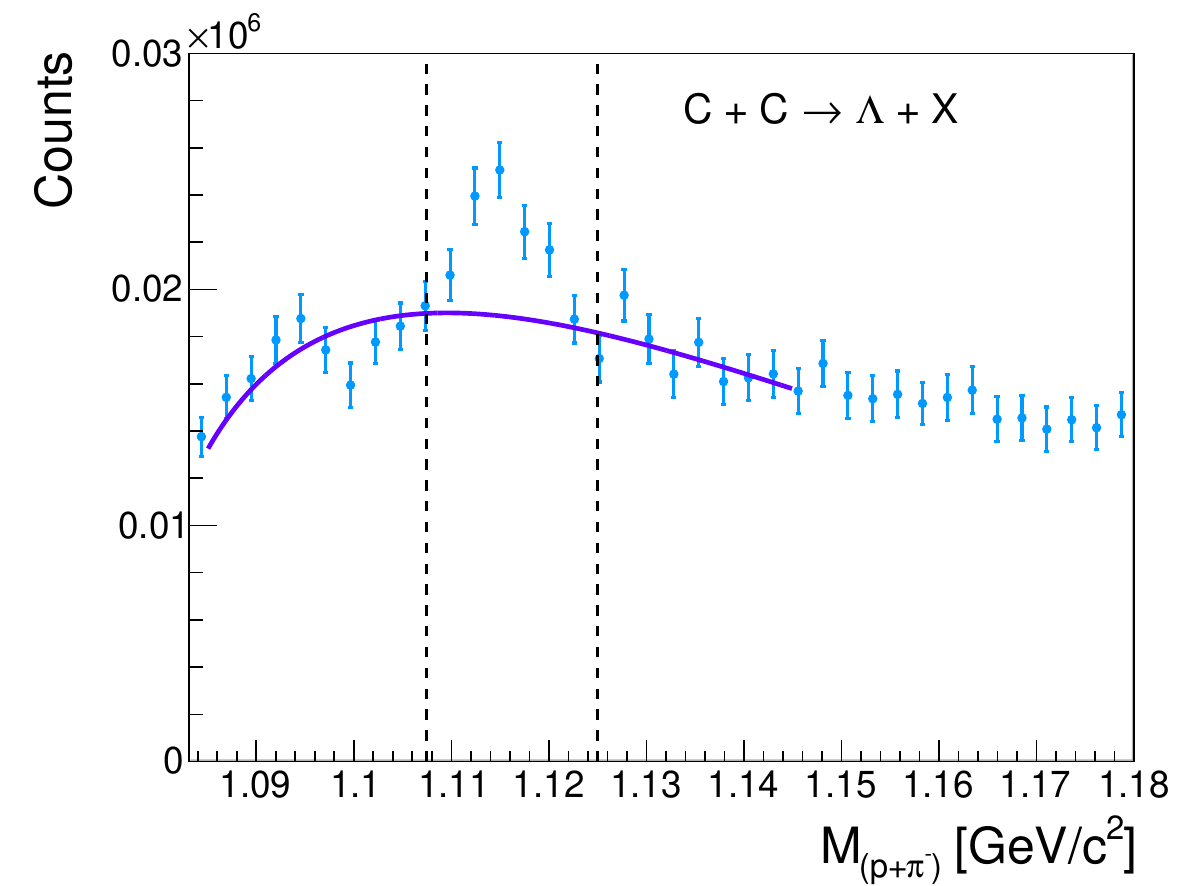}
\includegraphics[width=0.45\textwidth]{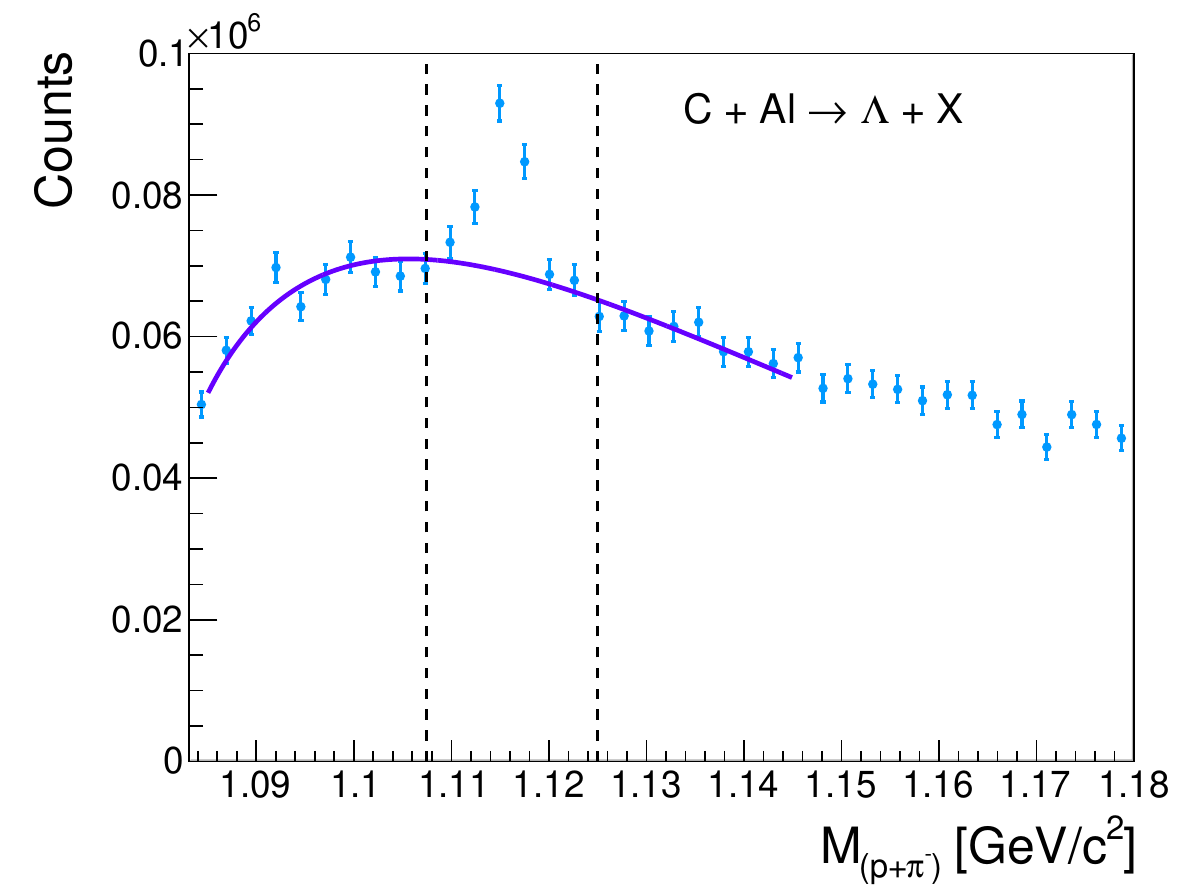}
\includegraphics[width=0.45\textwidth]{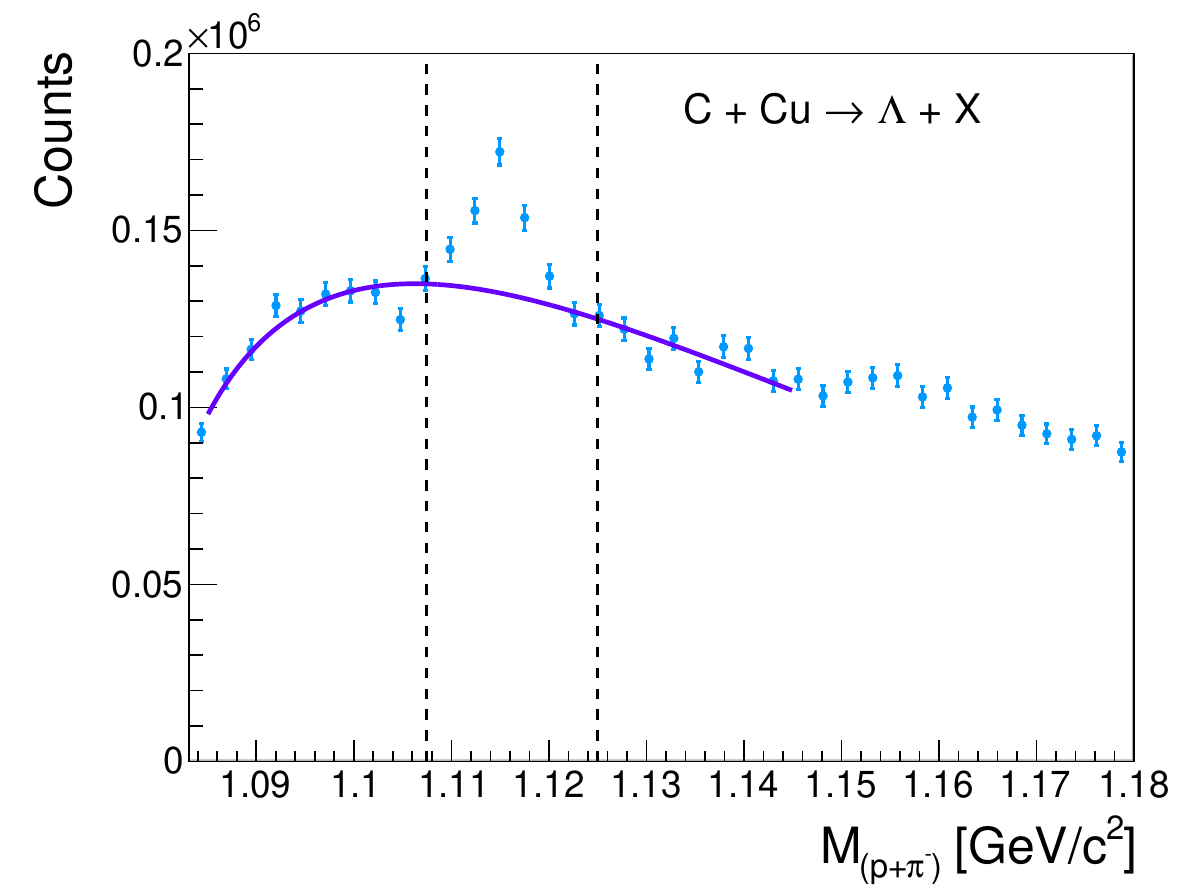}
\includegraphics[width=0.45\textwidth]{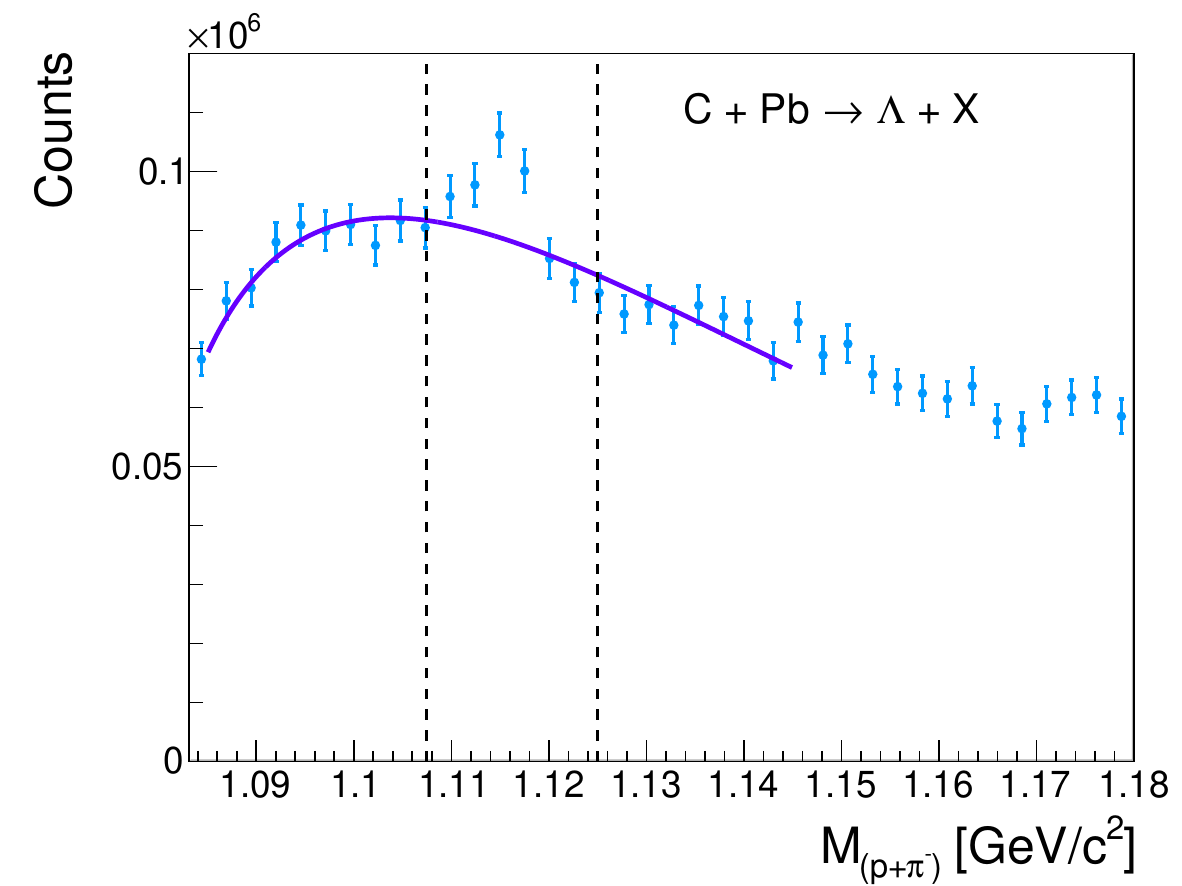}
 \end{center}
 \vspace{-0.2cm}
 \caption{Invariant mass spectra of $(p,\pi^-)$ pairs reconstructed in 
 interactions of the 4.5A~GeV carbon beam  with C, Al, Cu, and Pb targets. 
 The violet solid lines represent the result of the fit  according to Eq.~(2). 
 The vertical dashed lines show the mass window in which the $\Lambda$ 
 signal is calculated as the excess of the histogram relative to the background.}
\label{lamb}
\end{figure}

The number of $(p,\pi^-)$ pairs reconstructed in each ($y$, $p_T$) cell was multiplied 
by the inverse acceptance value for this cell. The resulting invariant mass distributions 
of  $(p, \pi^-)$ pairs produced in interactions of a 4.5A GeV carbon beam with C, Al, Cu, 
and Pb targets are shown in Figure~\ref{lamb}.

The background part of these distributions was parameterized according 
to Eq.~(2) in the 1.085--1.145~GeV/$c^2$ mass region.  
The region of the $\Lambda$ signal within the limits 1.1075--1.125~GeV/$c^2$
was excluded from the fit. 
The efficiency-corrected number of $\Lambda$ hyperons was calculated as the
sum of the bin contents within the mass-peak region after background
subtraction. The uncertainty of the background evaluation was 
propagated to the $\Lambda$ signal. The systematic uncertainty of the $\Lambda$ signal 
in data was calculated using the same bootstrap sampling method applied previously 
to Monte Carlo data (see Section~\ref{sect5}).


\section{Evaluation of $\Lambda$ hyperon cross sections and yields}
\label{sect7}

The inclusive cross section $\sigma_{\Lambda}$ and yield $Y_{\Lambda}$ of $\Lambda$ 
hyperon production in C~+~C, C~+~Al, C~+~Cu, and C~+~Pb interactions were
  calculated in $(y,~p_T)$ intervals according to the following formulae:

\begin{equation}
\sigma_{\Lambda}(y) = \sum_{p_T} \left[ \frac{N_{\Lambda}(y,p_T)}
{ \varepsilon_{\mathrm{trig}} \cdot L} \right], \quad
Y_{\Lambda}(y) = \frac{\sigma_{\Lambda}(y)}{\sigma_{\text{inel}}},
\end{equation}

\begin{equation}
\sigma_{\Lambda}(p_T) = \sum_{y} \left[ \frac{N_{\Lambda}(y,p_T)}
{\varepsilon_{\mathrm{trig}} \cdot L} \right], \quad
Y_{\Lambda}(p_T) = \frac{\sigma_{\Lambda}(p_T)}{\sigma_{\text{inel}}},
\end{equation}

\noindent where $L$ is the luminosity, $N_{\Lambda}$ is the number of acceptance 
corrected $\Lambda$ hyperons, $\varepsilon_{\mathrm{trig}}$ is the trigger efficiency, 
and $\sigma_{inel}$ is  the inelastic C~+~A interaction cross section. 
The luminosity was evaluated according to the method presented in
\cite{The_BM@N_collab_kaons_pions}.

The inelastic cross section for C~+~C interactions was taken from the measurement 
published in \cite{AngelovCC}. The inelastic cross sections for C~+~Al, C~+~Cu, 
and C~+~Pb interactions were calculated using the DCM-SMM model. The cross sections 
measured at various experiments have some scatter and can be parameterized by the 
formula ~\cite{Bradt1950, Barashenkov1966, Aksinenko1980}.
 
\begin{equation}
\sigma_{inel} = \pi R_0^2 \left(A_P^{1/3} + A_T^{1/3} - b\right)^2,
\end{equation}

\noindent with the parameters $R_0 = (1.46 \pm 0.01)$~fm, $b = (1.21 \pm 0.03)$ ~fm, 
at $\chi^2 = 3.5/NDF$, given in  \cite{AngelovCC}. Deviations of model calculations 
from the cross-section parametrization provided by~ Eq.\,(5) amount up to (10–12)\%. 

The values and uncertainties of $\sigma_{inel}$ for C~+~C, C~+~Al, C~+~Cu, and C~+~Pb 
interactions, used to evaluate $\Lambda$ hyperon yields, are presented in Table~\ref{L0prod}. 

The yields of $\Lambda$ hyperons in minimum bias C~+~C, C~+~Al, C~+~Cu, and C~+~Pb
interactions were measured in the kinematic range of the $\Lambda$ hyperon 
transverse momentum  $0.1<p_T<1.05$~GeV/c and the $\Lambda$ hyperon
rapidity in the laboratory frame $1.2<y<2.1$ for 4.0A~GeV and 4.5A~GeV data. 

The uncertainties of tracking reconstruction
efficiency, luminosity, and trigger are taken into account as part of the uncertainty of the weighted event.

For each $({y, ~p_{T})}$ interval, the uncertainty $\delta Y_{\Lambda}$
of the $\Lambda$ yield includes several sources:
\begin{itemize}
\item statistical fluctuations of the signal in experimental and MC 
      data -- $\delta Y_{\Lambda}^{stat}(data,mc)$;
\item systematic variations of the signal evaluated by the bootstrapping method 
      for experimental and MC data -- $\delta Y_{\Lambda}^{syst}(data,mc)$;
\item systematic uncertainties originating from variations of the $path$ and $dca$  
      within 10\% of the nominal value used in the analysis -- $\delta Y_{\Lambda}^{cuts}(data,mc)$.
\end{itemize}

\vspace{0.4cm}
\begin{equation}
\hspace{0pt}
\delta Y_{\Lambda} = \sqrt{
        \Bigl( \delta Y_{\Lambda}^{stat} (data,mc) \Bigr)^2 + 
	\Bigl( \delta Y_{\Lambda}^{syst} (data,mc) \Bigr)^2 +
        \Bigl( \delta Y_{\Lambda}^{cuts} (data,mc) \Bigr)^2}
\end{equation}

\par The corresponding values are shown in Tables~\ref{L0prod}, \ref{yield_y_table40}, and \ref{yield_pt_table40} of Section~8.  


\section{Results}
\label{sect8}

\begin{table}[htbp]
\begin{center}
  \caption{\small {$\Lambda$ hyperon production cross sections and yields at 4.0A~GeV 
  and 4.5A~GeV in C~+~C, C~+~Al, C~+~Cu, and C~+~Pb interactions. The first uncertainty 
  is statistical; the second one is systematic.}}
\begin{scriptsize}
\vspace{0.2cm}
\begingroup
\setlength{\tabcolsep}{5pt} 
\renewcommand{\arraystretch}{1.5} 
\begin{tabular}{|l|c|c|c|c|c|}
\hline
\multicolumn{2}{|l|}  { }                  & C + C                           & C + Al                         & C + Cu                         & C + Pb \\ \hline
\multicolumn{2}{|l|}{$\sigma_{inel}$, mb}        & $830 \pm 50$ \cite{AngelovCC} & $1250 \pm 50$ \cite{AngelovCC} & $1790 \pm 50$ \cite{AngelovCC} & $3075 \pm 120$ \cite{AngelovCC} \\\hline
Extrapolation factor                  &4.0A GeV  & $2.49 \pm 0.18$         & $3.01 \pm 0.13$        & $4.00 \pm 0.06$       &  $6.72 \pm 0.44$             \\ \cline{2-6}
to $4\pi$,  average                   &4.5A GeV  & $2.34\pm 0.08$          & $2.88 \pm 0.16$        & $3.76 \pm 0.15$       &  $6.24 \pm 0.14$             \\  \hline \hline
$\Lambda$ cross section,              &4.0A GeV  & $47.3 \pm 5.8 \pm 8.3$  & $121.0 \pm 15. \pm 31.$& $215. \pm 22. \pm 36.$ &   low statistics                \\
 extrapolated to $4\pi$, mb           &4.5A GeV  & $52.5 \pm 9.7 \pm 11.6$ &  $91. \pm 11.3 \pm 31.3$&   $249.0 \pm 36. \pm 40.$  & $633. \pm 192. \pm 192.$      \\ \hline \hline
$Y_{\Lambda} { /10^{-2}}$             &4.0A GeV  & $2.3 \pm 0.3 \pm 0.5$   & $3.2 \pm 0.4 \pm 0.8$  & $3.0 \pm 0.3 \pm 0.5$  &   low statistics           \\          \cline{2-6}
                                      &4.5A GeV  & $2.7 \pm 0.5 \pm 0.6$   & $2.5 \pm 0.3 \pm 0.7$  & $3.7 \pm 0.4 \pm 0.6$  & $3.3 \pm 0.1 \pm 0.1$         \\  \hline \hline
$Y_{\Lambda}^{4\pi} { /10^{-2}}$      &4.0A GeV  & $5.7 \pm 0.7 \pm 1.0$   & $9.6 \pm 1.0 \pm 2.5$  & $12.0 \pm 1.0 \pm 2.0$ &   low statistics           \\           \cline{2-6}
                                      &4.5A GeV  & $6.3 \pm 1.2 \pm 1.4 $  & $7.1 \pm 0.9 \pm 2.5$     & $14.0 \pm 2.0 \pm 2.0$     & $ 20.0 \pm 6.0 \pm 6.0$       \\ \hline \hline
                                                           \multicolumn{6}{|c|}{Number of participants, model prediction}  \\  \hline
$N_{part}$, {DCM-SMM}                 &     &   9.0                   &  13.4                  &  23.0                  &  50.5                      \\
$N_{part}$, {UrQMD}                   &     &   7.2                   &  11.4                  &  19.3                  &  50.0                      \\
$N_{part}$, {PHSD}                    &     &   8.4                   &  11.9                  &  17.3                  &  30.8                      \\   \hline \hline
$Y_{\Lambda}^{4\pi}/N_{part}^{DCM-SMM} {/10^{-3}}$&4.0A GeV   & $6.3 \pm 0.08 \pm 0.1$  & $7.2 \pm 0.8 \pm 2.0$  & $5.7 \pm 0.4 \pm 0.3$  &  low statistics \\                      \cline{2-6}
                                      &4.5A GeV      & $7.0 \pm 0.1 \pm 0.2 $  & $5.3 \pm 0.7 \pm 1.9$     & $6.1 \pm 0.9 \pm 0.9$      & $ 3.8 \pm 1.1 \pm 1.1$        \\  \hline
\end{tabular}
\endgroup
\end{scriptsize}
\label{L0prod}
 \end{center}
\end{table}
The production cross sections and yields of $\Lambda$ hyperons, integrated
luminosity, and extrapolation factors to the full phase space for C~+~C, C~+~Al, C~+~Cu, and C~+~Pb
interactions are shown in Table~\ref{L0prod}. 
\begin{figure}[htbp]
\begin{center}
\includegraphics[width=1.0\textwidth]{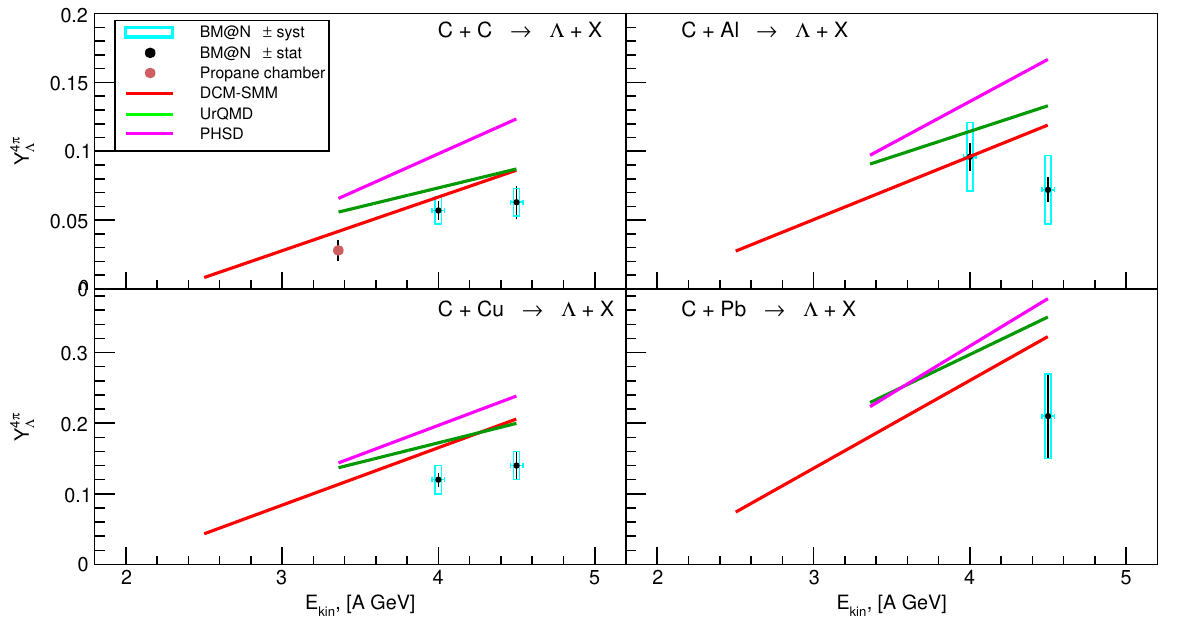}
 \end{center}
\vspace{-0.5cm}
 \caption{\small Energy dependence of $\Lambda$ hyperon $Y^{4\pi}_{\Lambda}$ yields 
 in C~+~C,  C~+~Al, C~+~Cu, and C~+~Pb interactions. 
 The result of the Propane Chamber experiment~\cite{ArakelianCC,ArmutCC}  for C~+~C is 
 shown for comparison.  The error bars represent the statistical uncertainties; 
 the blue boxes show the systematic uncertainties. The predictions of the 
 DCM-SMM, UrQMD, and PHSD models are shown as colored lines.
 }
\label{yields_energy40}
\end{figure}
\begin{figure}[htbp]
\centering
\includegraphics[width=\textwidth]{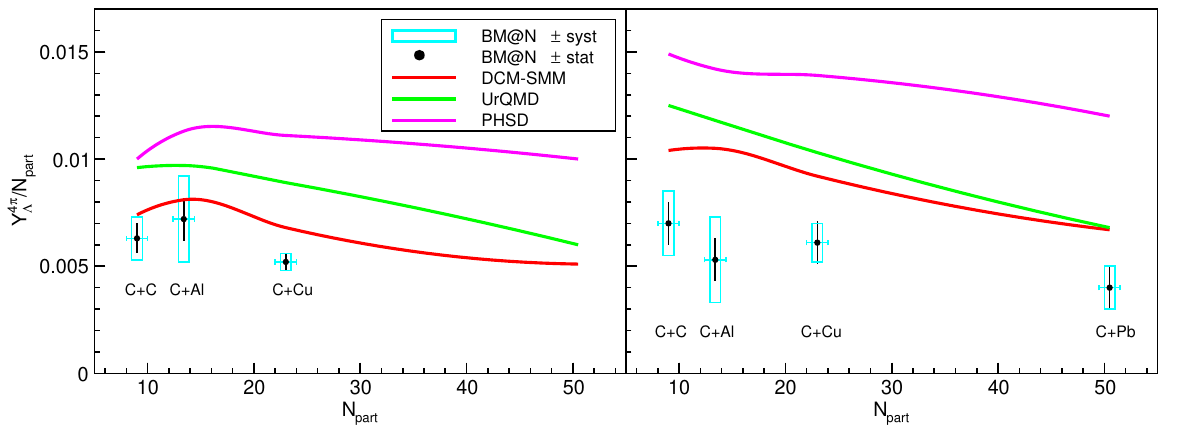}
\caption{{\small 
    Ratios of the $\Lambda$ hyperon yields to the number of nucleon participants measured by BM@N 
    in minimum bias carbon--nucleus interactions at 4.0A~GeV (left) and 4.5A~GeV (right) 
    compared with the model predictions. } }
    \label{fig10}
\end{figure}
The extrapolation factors were calculated by averaging the predictions of 
the DCM-SMM~\cite{DCM_QGSM,DCM_SMM} and UrQMD~\cite{UrQMD} models. 
In nucleus-nucleus collisions with different atomic masses, the 
conventional quantity used in data analysis is the number of participants ($N_{part}$).
The number of participants was calculated for C~+~C, C~+~Al, C~+~Cu, and C~+~Pb 
interactions on the basis of the DCM-SMM, UrQMD, and PHSD~\cite{PHSD} models. 
The $N_{part}$ values are the same for 4.0A GeV and 4.5A GeV.
The yields of $\Lambda$ hyperons normalized to $N_{part}$ from the DCM-SMM are shown in Table~\ref{L0prod}. The measured $\Lambda$ yields ($Y_{\Lambda}$) within the BM@N acceptance were 
extrapolated to the full kinematic range ($Y^{4\pi}_{\Lambda}$). 

The $Y^{4\pi}_{\Lambda}$ yields as a function of 
the carbon beam kinetic energy ($E_{kin}$) are shown in Figure~\ref{yields_energy40}.
The result of the Propane Chamber experiment~\cite{ArakelianCC,ArmutCC} on the $\Lambda$ yield
in C~+~C interactions with a kinetic energy of 3.36A~GeV is shown in Figure~\ref{yields_energy40} 
(top left) for comparison. 
The BM@N results are compared with the 
 predictions of the DCM-SMM, UrQMD, and PHSD models. 
Models predict the increase of the $\Lambda$ hyperon yield with increasing beam energy. All models predict higher yields than was measured and exceed the experimental data. The better agreement was achieved with the DCM-SMM model, while the PHSD model predicts considerably higher total $\Lambda$ hyperon yield.

\begin{table}[htbp]
  \caption{\small {$\Lambda$ hyperon yields as a function of rapidity 
   for C~+~C, C~+~Al, C~+~Cu, and C~+~Pb interactions. The first uncertainty is statistical; 
   the second one is systematic.}}
\begin{scriptsize}
\vspace{0.3cm}
\begingroup
\setlength{\tabcolsep}{5pt} 
\renewcommand{\arraystretch}{1.5} 

\begin{tabular}{|l|c|c|c|c|c|}
\hline
\multicolumn{2} {|l|}  { }             & C + C                          & C + Al                          & C + Cu                      & C + Pb \\ \hline
$1.20<y<1.45$  &4.0A~GeV     & $0.036 \pm 0.008 \pm 0.003 $ & $0.064 \pm 0.009 \pm 0.005 $ & $0.042 \pm 0.009\pm 0.005 $ &   low statistics            \\ \cline{2-6}
               &4.5A~GeV     & $0.051 \pm 0.015 \pm 0.008 $ & $0.042 \pm 0.009 \pm 0.005 $ & $0.058 \pm 0.014\pm 0.005 $ & $ 0.051\pm 0.02\pm 0.008 $  \\ \hline \hline
$1.45<y<1.65$  &4.0A~GeV     & $0.022 \pm 0.005 \pm 0.002 $ & $0.046 \pm 0.007 \pm 0.004 $ & $0.047 \pm 0.007\pm 0.004 $ &   low statistics            \\ \cline{2-6}
               &4.5A~GeV     & $0.027 \pm 0.008 \pm 0.008 $ & $0.044 \pm 0.006 \pm 0.004 $ & $0.051 \pm 0.007\pm 0.004 $ & $ 0.039\pm 0.01\pm 0.005 $  \\ \hline \hline
$1.65<y<1.85$  &4.0A~GeV     & $0.021 \pm 0.004 \pm 0.002 $ & $0.019 \pm 0.005 \pm 0.004 $ & $0.033 \pm 0.005\pm 0.004 $ &   low statistics            \\ \cline{2-6}
               &4.5A~GeV     & $0.017 \pm 0.008 \pm 0.008 $ & $0.021 \pm 0.005 \pm 0.004 $ & $0.039 \pm 0.006\pm 0.004 $ & $ 0.034\pm 0.009\pm 0.004 $ \\ \hline \hline
$1.85<y<2.10$  &4.0A~GeV     & $0.012 \pm 0.004 \pm 0.002 $ & $0.007 \pm 0.005 \pm 0.004 $ & $0.024 \pm 0.005\pm 0.004 $ &   low statistics            \\ \cline{2-6}
               &4.5A~GeV     & $0.033 \pm 0.008 \pm 0.008 $ & $0.015 \pm 0.004 \pm 0.004 $ & $0.031 \pm 0.005\pm 0.004 $ & $ 0.012\pm 0.007\pm 0.004 $ \\ \hline
\end{tabular}
\endgroup
\end{scriptsize}
\label{yield_y_table40}
\end{table}

A comparison of the measured 4$\pi$ extrapolated yields
$Y^{4\pi}_{\Lambda}$ normalized to the number of participants $N_{part}$ with the 
predictions of the DCM-SMM, UrQMD, and PHSD models 
for 4.0A~GeV and 4.5A~GeV carbon--nucleus interactions is shown in Figure~\ref{fig10}.
The measured integral yields per participant show either independence of the number 
of participants or some decrease for heavier target nuclei as predicted by the theoretical models.

\begin{figure}[htbp]
\begin{tabular}{lr}
\centering
\begin{minipage}[l]{6.5cm}
\vspace*{-0.5cm}
\includegraphics[width=1.05\textwidth]{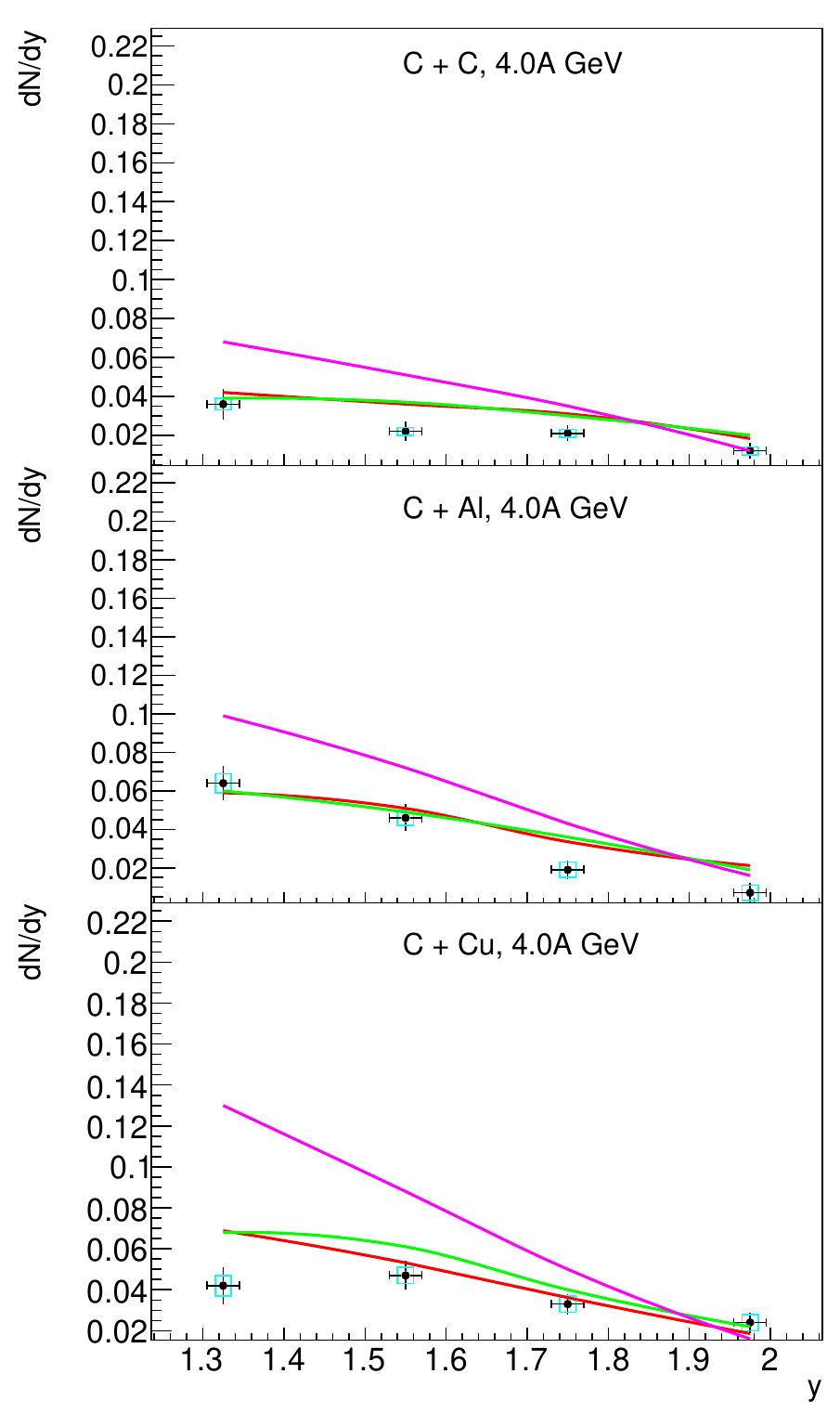}
{\hspace*{0.6cm}
\includegraphics[width=0.8\textwidth]{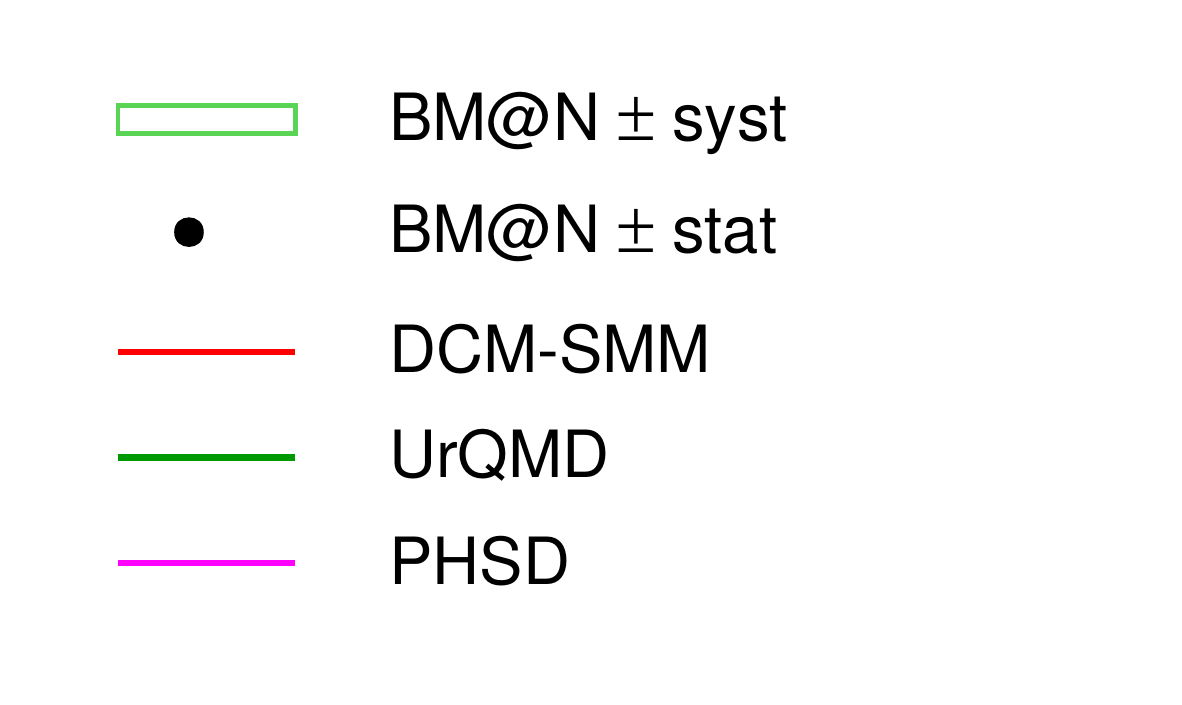}
}
\end{minipage}
&
\begin{minipage}[r]{6.5cm}
\vspace{0.2cm}
\includegraphics[width=1.05\textwidth]{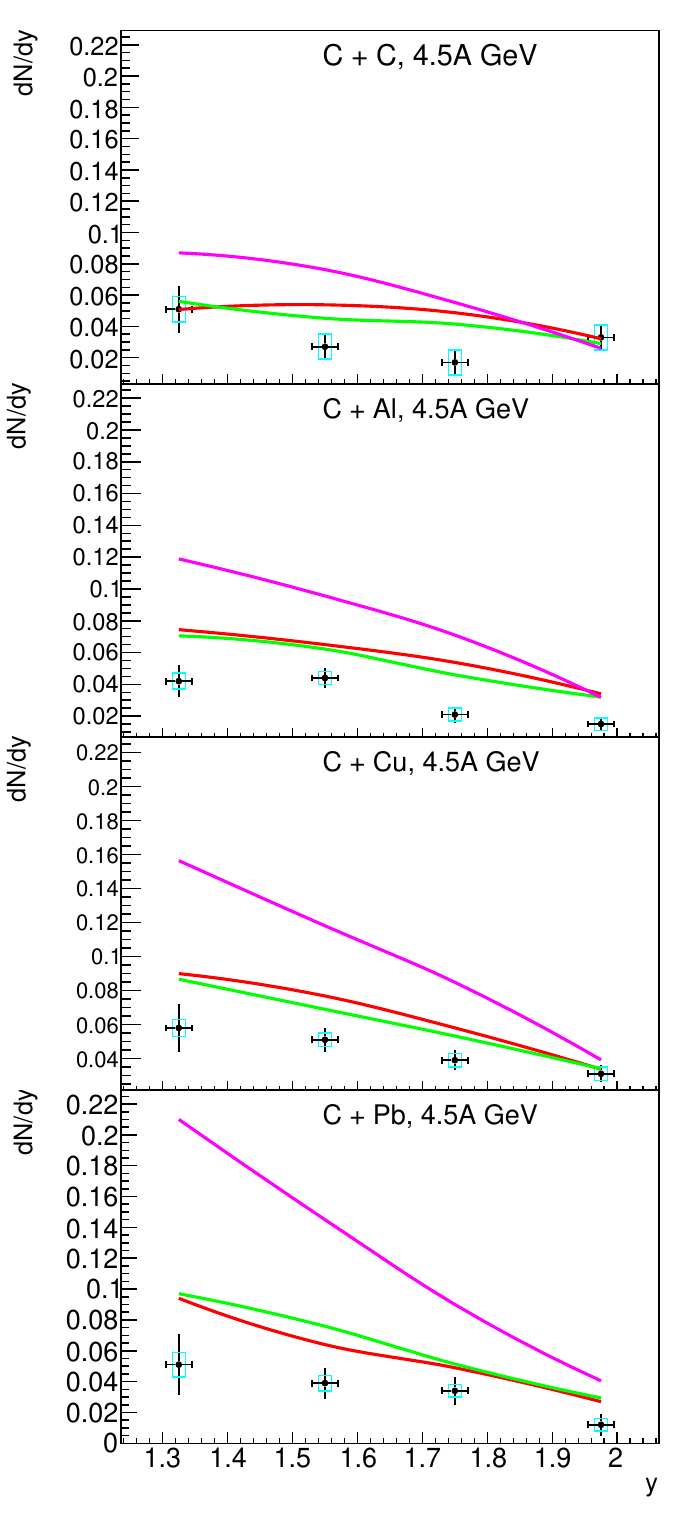}
\end{minipage}
\end{tabular}
 \caption{\small Rapidity distributions of $\Lambda$ hyperons produced 
 in C~+~C, C~+~Al, and C~+~Cu 
 in\-te\-rac\-tions at the car\-bon beam ener\-gy of 4.0A~GeV (left column) 
 and 4.5A~GeV (right column).}
\label{yields_y}
\end{figure}

The rapidity distributions of $\Lambda$ hyperons were measured in the transverse momentum 
range of $0.1<p_T<1.05$~GeV/c. 
The differential $y$ rapidity spectra of $\Lambda$ hyperons in the laboratory frame, 
corrected for the detector acceptance and efficiency, are presented in Figure~\ref{yields_y} 
and the corresponding numerical values are listed in Table~\ref{yield_y_table40}.
The transformation of the $y$ distribution from the laboratory system to the center-of-mass system 
is given by a shift $y_{CM}=y-1.17(1.22)$ for 4.0~(4.5)A~GeV.

\begin{figure}[htbp]
\begin{tabular}{lr}
\begin{minipage}[l]{6.5cm}
\vspace*{-0.5cm}
\centering
\includegraphics[width=1.05\textwidth]{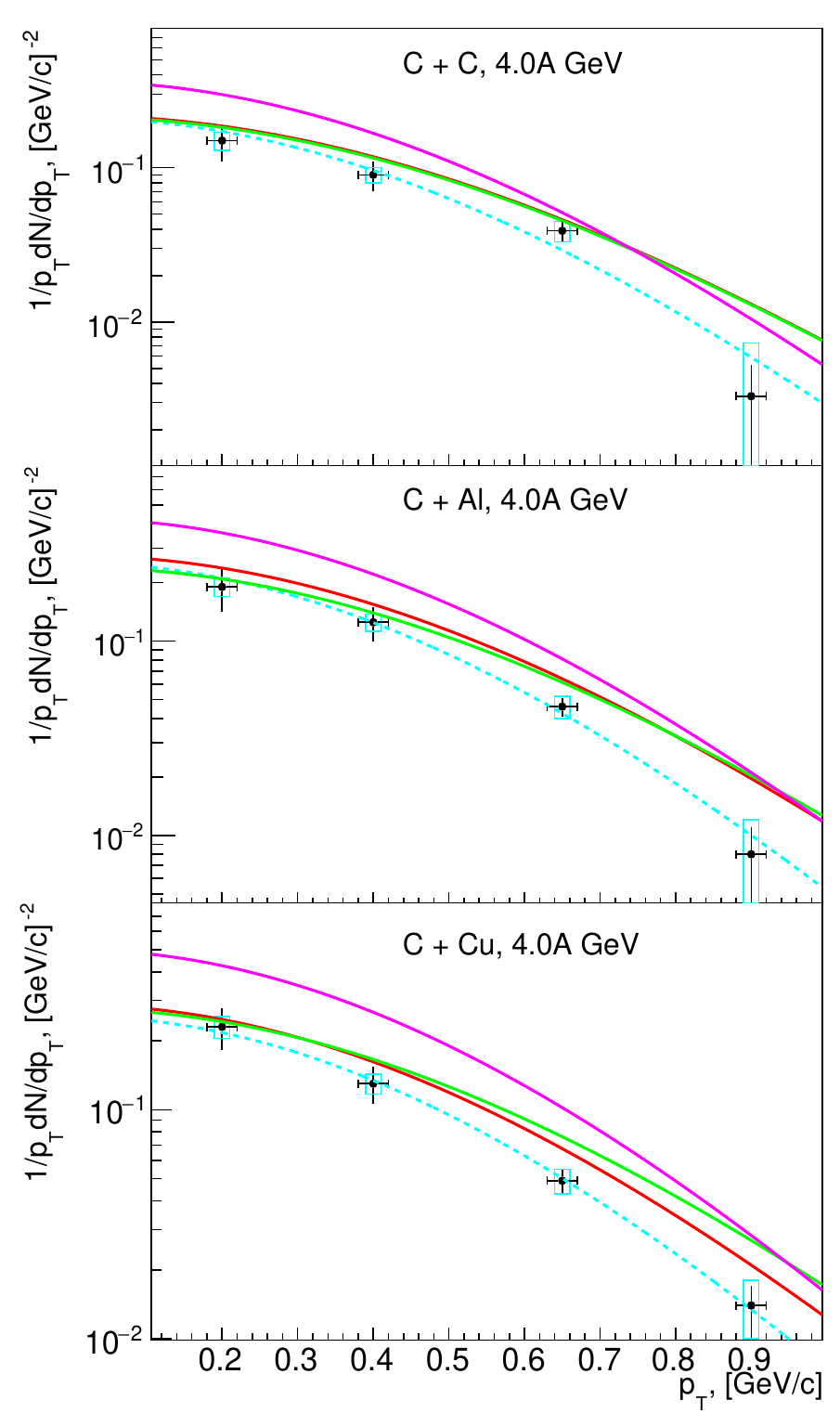}
{\hspace*{0.6cm}
\includegraphics[width=0.8\textwidth]{pt_symbols.pdf}
}
\end{minipage}
&
\begin{minipage}[r]{6.5cm}
\vspace{0.2cm}
\includegraphics[width=1.05\textwidth]{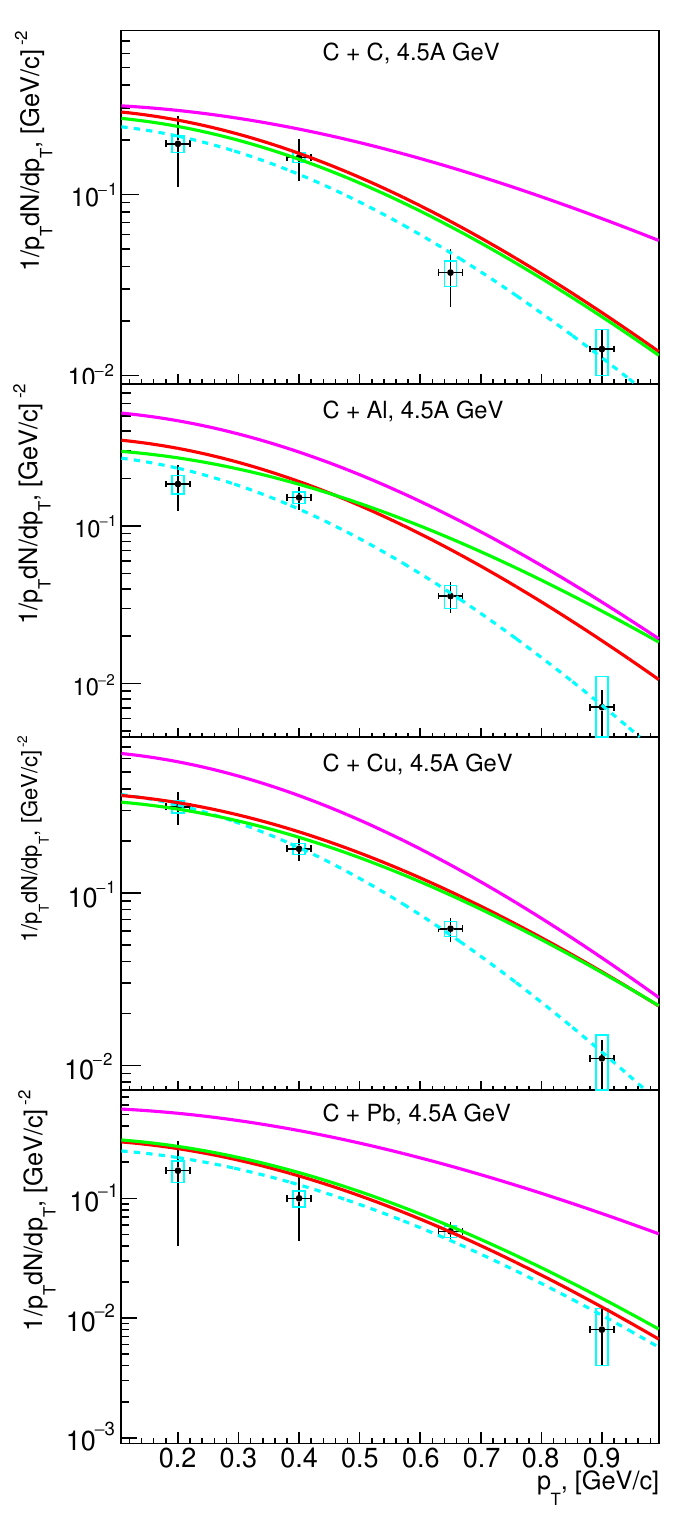}
\end{minipage}
\end{tabular}
 \caption{\small {Transverse momentum distributions of $\Lambda$ hyperons produced in interactions of the
 carbon beam with C~+~C, C~+~Al, and C~+~Cu targets at energies of 4.0A~GeV (left plots)
and 4.5A~GeV (right plots).
The blue lines represent the results of the parameterization described in the text. }}
 \label{yields_pt}
\end{figure}

The  predictions of the DCM-SMM, UrQMD, and PHSD models are shown in Figure~\ref{yields_y} 
and Figure~\ref{yields_pt} for comparison. The models describe the shape of the differential spectra 
in $y$ and $p_T$ in general. However, in most cases, they overestimate the measured $\Lambda$ yields.
The predictions of the DCM-SMM and UrQMD models are closer to experimental
results than those of the PHSD model. The PHSD model predicts a stronger rapidity dependence 
of the $\Lambda$ hyperon yield compared to the DCM-SMM and UrQMD models and provides worse 
agreement with BM@N data.
 
The transverse momentum spectra of $\Lambda$ hyperons are presented in 
Figure~\ref{yields_pt} and Table~\ref{yield_pt_table40}. They were parameterized in the 
measured $y$ range by the function:
\begin{equation}
\frac{1}{p_T} \cdot \frac{d^2N}{dp_T\,dy} \propto \exp\left(-\frac{m_T - m_{\Lambda}}{T_0}\right),
\end{equation}
\noindent where $m_T=\sqrt{m_{\Lambda}^2+p_T^2}$ is the transverse 
mass, the  inverse slope parameter $T_0$ is a free parameter of the fit.

The inverse slope $T_0$ values resulting from the fit of the invariant $p_T$ spectra 
are shown in Table~\ref{Tslopes} for 4.0A~GeV and 4.5A~GeV carbon beam data, respectively. The calculated numbers show an increase in the  $T_0$ inverse slope parameter with increasing target mass. Although statistical and systematic uncertainties do not allow a clear conclusion regarding the dependence of the spectral slopes on the system-size or energy.

\begin{table}[!hbp]
\caption{{\small $\Lambda$ hyperon yields as a function of transverse momentum 
          for C~+~C, C~+~Al, C~+~Cu, and C~+~Pb interactions. The first uncertainty 
	  is statistical; the second one is systematic.}}
\begin{scriptsize}
\vspace{0.3cm}
\begingroup
\setlength{\tabcolsep}{5pt} 
\renewcommand{\arraystretch}{1.5} 

\begin{tabular}{|l|c|c|c|c|c|}
\hline
\multicolumn{2} {|l|}  { }            & C~+~C                         & C~+~Al                          & C~+~Cu                   & C~+~Pb  \\  \hline
$0.10<p_T<0.30$  &4.0A~GeV     & $0.031 \pm 0.004\pm 0.003 $ & $0.039 \pm 0.009\pm 0.004 $ & $0.047 \pm 0.009\pm 0.005 $ &   low statistics                \\  \cline{2-6}
                 &4.5A~GeV     & $0.027 \pm 0.008\pm 0.003 $ & $ 0.037\pm 0.009\pm 0.005 $ & $0.06  \pm 0.01 \pm 0.005 $ & $0.03  \pm 0.02\pm 0.007 $      \\  \hline \hline
$0.30<p_T<0.50$  &4.0A~GeV     & $0.036 \pm 0.004\pm 0.003 $ & $0.063 \pm 0.009\pm 0.005 $ & $0.054 \pm 0.007\pm 0.005 $ &   low statistics                \\  \cline{2-6}
                 &4.5A~GeV     & $0.064 \pm 0.006\pm 0.004 $ & $0.060 \pm 0.004\pm 0.005 $ & $0.07\   \pm 0.01\  \pm 0.005 $ & $0.06  \pm 0.01 \pm 0.006 $ \\ \hline \hline
$0.50<p_T<0.75$  &4.0A~GeV     & $0.025 \pm 0.002\pm 0.003 $ & $0.030 \pm 0.005\pm 0.004 $ & $0.032 \pm 0.005\pm 0.004 $ &   low statistics                \\  \cline{2-6}
                 &4.5A~GeV     & $0.025 \pm 0.003\pm 0.003 $ & $0.023 \pm 0.003\pm 0.004 $ & $0.041 \pm 0.006\pm 0.004 $ & $0.034 \pm 0.007\pm 0.004 $     \\  \hline \hline
$0.75<p_T<1.05$  &4.0A~GeV     & $0.004 \pm 0.001\pm 0.001 $ & $0.013 \pm 0.003\pm 0.004 $ & $0.013 \pm 0.005\pm 0.004 $ &   low statistics                \\  \cline{2-6}
                 &4.5A~GeV     & $0.012 \pm 0.001\pm 0.002 $ & $0.006 \pm 0.003\pm 0.002 $ & $0.011 \pm 0.002\pm 0.004 $ & $0.007 \pm 0.004\pm 0.002 $     \\ \hline
\end{tabular}
\endgroup
\end{scriptsize}
\label{yield_pt_table40}
\end{table}

The fully integrated $4\pi$ $\Lambda$ yields $(n_{CC}^{\Lambda})$ in C~+~C collisions 
were compared with a calculation based on the parametrization of the total $\Lambda$ 
hyperon yield $n_{pp}^\Lambda$ in proton--proton ($p+p$) 
interactions  ~\cite{W.Cassing, V.Kolesnikov} and scaled to the C~+~C system by the 
number of participants and corrected for isospin effects.

The parametrization is based on the Lund String Model (LSM) ~\cite{W.Cassing} and expressed as:
\begin{equation}
 n_{pp}^\Lambda  = a(x - 1)^b x^{-c},
\end{equation}

\noindent where $x = {s}/{s_0}$ is the ratio of the center-of-mass energy squared $s$ 
and the production threshold energy squared ${s_0}$, and $a$, $b$, $c$ are fit 
parameters~\cite{V.Kolesnikov}.

Since C~+~C collisions include not only $p+p$ but also $p+n$ and $n+n$ interactions, 
and taking into account that the near-threshold $\Lambda$ yields are about 25\% lower 
in $n+n$ and $n+p$ compared to $p+p$ collisions~\cite{V.Kireyeu}, the following isospin 
correction factor was applied:
\begin{equation}
k_{\text{iso}} = f_{pp} \cdot \alpha + (f_{np} + f_{pn} + f_{nn}) \cdot \beta,
\end{equation}

\noindent with $\alpha = 1.0$ for $p+p$ and $\beta = 0.75$ for $n+n$, $n+p$, 
and $p+n$ collisions. The fractions $f_{ij}$ are determined by the composition of 
nucleons in the colliding carbon nuclei and targets (e.g. $f_{np}=f_{pn}=f_{pp}=f_{nn}=0.25$ 
for $^{12}C+^{12}C$ collisions).

The total yield $Y_{\Lambda}^{4\pi}$ for C~+~C was scaled as:
\begin{equation}
 Y_{\Lambda}^{4\pi}  =  n_{pp}^\Lambda \cdot k_{\text{iso}} \cdot N_{\text{part}},
\end{equation}

\noindent where $N_{\text{part}}$ is the number of participating nucleons, 
and $k_{\text{iso}}$ is the isospin correction factor  in Eq. (9) for the $4\pi$ 
extrapolated $\Lambda$ yields.


\begin{table}[htbp]
\centering
  \caption{{\small The inverse slope parameters of the $p_T$ spectra for 
                      C~+~C, C~+~Al, C~+~Cu, and C~+~Pb configurations at 4.0--4.5A~GeV. 
                      The first value is statistical and the second one is systematic uncertainty.}}
\begin{scriptsize}
\vspace{0.3cm}
\begingroup
\setlength{\tabcolsep}{5pt} 
\renewcommand{\arraystretch}{1.5} 
\begin{tabular}{|l|c|c|c|c|c|}
\hline
\multicolumn{2}{|l|}{}         & C~+~C                        & C~+~Al             & C~+~Cu                  & C~+~Pb              \\ \hline
Inverse slope $T_0$,  MeV  & 4.0A~GeV    &  $89 \pm 9 \pm 17$   & $99 \pm 10 \pm 16$ & $108 \pm 11 \pm 14$ &   low statistics         \\ \cline{2-6}
                           & 4.5A~GeV    &  $107 \pm 17 \pm 17$ & $86 \pm  8 \pm 17$ & $ 91 \pm 8  \pm 15 $ & $99 \pm 17 \pm 20$    \\
\hline
\end{tabular}
\endgroup
\end{scriptsize}
\label{Tslopes}
\end{table}

\begin{figure}[htb]
\centering
\includegraphics[width=0.6\textwidth]{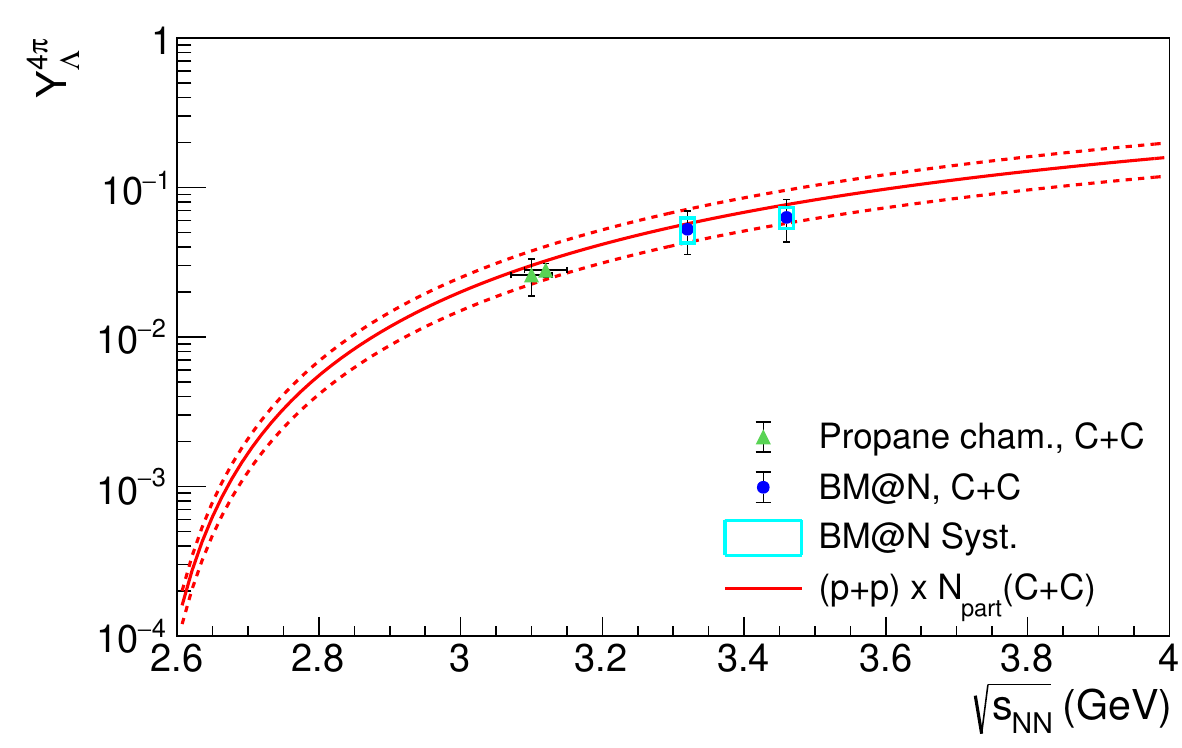}
\caption{\small  The integrated yield of $\Lambda$ hyperons in C~+~C collisions as a function 
 of $\sqrt{s_{NN}}$. BM@N experimental data are compared with a parameterization 
 based on $pp$ collisions     scaled to $N_{\text{part}}=9$. Dashed red lines 
 indicate the uncertainties in the predicted excitation function (about 25\%). } 
\label{snn}
\end{figure}
The BM@N results for $\Lambda$ yields in C~+~C collisions at 4.0A~GeV and 4.5A~GeV are in 
good agreement with the scaled $p+p$ parameterization model as shown in Figure~\ref{snn}.
The parameterization provides a reliable basis for estimating $\Lambda$ hyperon 
production in carbon--carbon interactions. 
The agreement with the BM@N experimental data supports its applicability for light 
symmetric systems. In addition, it provides an independent cross-check of the number of 
participating nucleons $N_{\text{part}}$ used in the scaling, which was taken from 
the DCM-SMM model and evaluated according to existing measurements from the 
Propane Chamber experiment. 

\begin{figure}[htb]
\vspace{0.6cm}
\begin{tabular}{lr}
\begin{minipage}{0.49\textwidth}
        \includegraphics[width=1.02\textwidth]{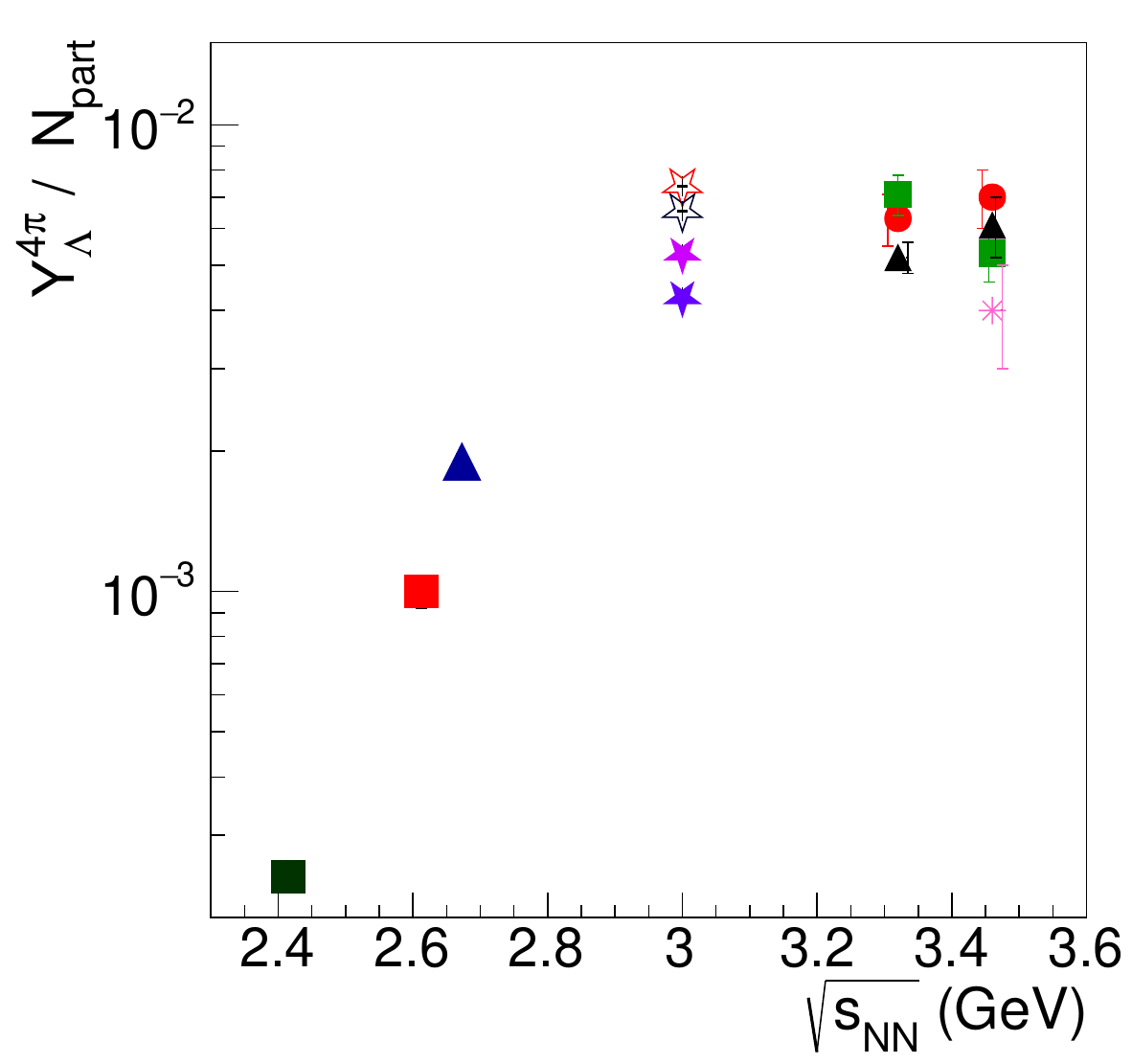}
\end{minipage}
&
\begin{minipage}{0.49\textwidth}

\begin{scriptsize}
\begingroup
\setlength{\tabcolsep}{5pt}       
\renewcommand{\arraystretch}{1.5} 
\begin{tabular}{|l|c|c|c|c|}
\hline
Experiment      &        \multicolumn{2}{|c|}{beam+target}       &$\sqrt{s_{NN}}, GeV$ & $N_{part}$    \\ \hline
BM@N            & {\color{red}   $\bullet$}            & C+C     & 3.32               & \ 9.0             \\ 
4.0A GeV        & {\color{Green}    $\blacksquare$}       & C+Al    &            &  13.4             \\
                & {\color{black}    $\blacktriangle$}     & C+Cu    &                &  23.0             \\ \hline
BM@N            & {\color{red}      $\bullet$}            & C+C     &  3.46              & \ 9.0              \\ 
4.5A GeV        & {\color{Green} $\blacksquare$}      & C+Al    &            &  13.4             \\
                & {\color{black}     $\blacktriangle$}      & C+Cu    &                &  23.0             \\
                & {\color{magenta} $\ast$}      & C+Pb    &                &  50.5             \\ \hline \hline
HADES           & {\color{black}    $\blacksquare$}       & Au+Au   &  2.4           &  193          \\ 
                & {\color{red}       $\blacksquare$}       & Ar+KCl  &  2.61          &  39           \\ \hline
FOPI            & {\color{Blue} $\blacktriangle$}     & Ni+Ni   &  2.67          &  71           \\ \hline
STAR            
                & {\color{Red3}     \FiveStarOpen}       & Au+Au   &  3.0           &  159.5           \\     
                & {\color{black}    \FiveStarOpen}       & Au+Au   &                &  110.5           \\   
                
                & {\color{Magenta3}     $\bigstar$}           & Au+Au   &                &  59           \\ 
                & {\color{blue}         $\bigstar$}           & Au+Au   &                &  20           \\ \hline
\end{tabular}
\endgroup
\end{scriptsize}
\end{minipage}
\end{tabular}

\caption{{\small Integral $\Lambda$ yields normalized
           to  $N_{\text{part}}$ as a function of
           collision energy $\sqrt{s_{NN}}$. }}

\label{fig:lambda_yield_all}
\end{figure}

The measured fully integrated $4\pi$ $\Lambda$ yields
were compared with the results from other heavy-ion experiments in Figure~\ref{fig:lambda_yield_all}. 
To allow for a direct comparison, all yields were normalized to the corresponding
average numbers of participating nucleons $N_{\text{part}}$, strongly depending on the 
colliding nuclei and collision centralities, see Table in Figure~\ref{fig:lambda_yield_all}.

A comparison of the normalized $\Lambda$ yields at lower energies, measured by the HADES collaboration 
in Au~+~Au at 1.23A~GeV ($\sqrt{s_{NN}}= 2.4$ GeV)~\cite{HADES_AuAu_2018} 
and Ar~+~KCl at 1.76A~GeV ($\sqrt{s_{NN}}=2.61$  GeV)~\cite{HADES2010}, as well as 
the FOPI result in Ni~+~Ni at 1.93A~GeV ($\sqrt{s_{NN}}=2.67$ GeV) \cite{FOPI_K0_Lambda}, 
shows a smooth increase in the yield with rising collision energy. For $\sqrt{s_{NN}}=3.0$ GeV, 
the integrated $\Lambda$ yields and corresponding values of
$N_{\text{part}}$ for each centrality class were taken from the publication by the 
STAR collaboration\,\cite{STAR_Strangeness}. 

The STAR results for relatively peripheral collisions with $N_{\text{part}} \simeq 20$  and $N_{part} \simeq 59$  are in agreement with the BM@N data points obtained at similar values of $N_{\text{part}}$. For more central Au~+~Au collisions, the STAR measurements  show higher normalized $\Lambda$ yields.

This comparison indicates that, after normalization by the number of
participating nucleons, $\Lambda$ production exhibits a smooth energy dependence across different collision systems within an overlapping $N_{\text{part}}$ range.


\section{Summary}
\label{sect9}

The BM@N experiment, exploring the advantages of the fixed-target setup, studied $\Lambda$ hyperon production in symmetric and asymmetric beam--target configurations. The data sets were collected with two kinetic energies of the carbon beam 4.0A~GeV and 4.5A~GeV, with one symmetric C~+~C and three asymmetric C~+~Al, C~+~Cu, and C~+~Pb layouts.
 These results are compared with the predictions of DCM-SMM, UrQMD, and PHSD models. For the C~+~C reaction, the $\Lambda$ hyperon yield is also compared with the results of the Propane Chamber experiment obtained in C~+~C interactions at lower energy.
The $\Lambda$ hyperon production cross sections were evaluated to be ($47.3 \pm 5.8$~mb) and ($52.5 \pm 9.7$~mb) for C~+~C collisions at energies of 4.0A~GeV and 4.5A~GeV, respectively. These values are about twice as large as the results of the Propane Chamber experiment at 3.36A~GeV ($24 \pm 6$~mb), indicating a general rise in the production cross section with increasing collision energy.
The cross sections and yields of $\Lambda$ hyperons in C~+~C, C~+~Al, C~+~Cu, and C~+~Pb (only at 4.5A~GeV)
collisions are presented in Table~\ref{L0prod}  for both beam energies, as well as in Figure~\ref{yields_energy40}. 

The BM@N results for $\Lambda$ production in C~+~C collisions at 4.0A~GeV and 4.5A~GeV show 
good agreement with a proton--proton-based parameterization model scaled to the carbon--carbon
system. The scaling takes into account the number of participants involved in the reaction, estimated by
the DCM-SMM model, as well as the isospin effects.

The $\Lambda$ yields from different experiments are scaled by the average number of participants, $N_{\text{part}}$, to allow a direct comparison between various colliding nuclear systems (Figure~\ref{fig:lambda_yield_all}). 
The results of this scaling are consistent with a general  energy dependence for various collision systems measured  over an overlapping $N_{\text{part}}$ range.

There is an indication of an increase in the slope parameter $T_0$, extracted from exponential fits to the transverse momentum spectra, with target mass. However, the statistical and systematic uncertainties do not allow a definitive conclusion regarding the system-size or the energy dependence of the spectral slopes.
A more precise determination of these dependencies will require additional statistics to
be collected in future BM@N runs.

The present BM@N results confirm the feasibility of high quality $\Lambda$ hyperon production studies with different beam–target configurations at Nuclotron facility.

\vspace{-0.5cm}
\paragraph{Acknowledgments:}

The BM$@$N Collaboration acknowledges the efforts of the staff of the accelerator division of the
Laboratory of High Energy Physics at JINR that made this experiment possible.
The BM$@$N Collaboration acknowledges support of the HybriLIT of JINR for the provided computational 
resources. The BM@N Collaboration gratefully acknowledges the support of the BM@N DAQ Cluster 
Team for providing the necessary resources and facilities that contributed to this research. 
The research has been supported by the Ministry of Science and Higher Education of the 
Russian Federation, Project ``New Phenomena in Particle Physics and the Early Universe'' 
No. FSWU-2023-0073. 
The work was funded by the Ministry of Science and Higher Education of
the Russian Federation, Project "Studying physical phenomena in the micro-
and macro-world to develop future technologies" FSWU-2026-0010
and by the Science Committee of the Ministry of Science and Higher Education of the 
Republic of Kazakhstan (Grant No. AP23487706).

\label{sect10}

\end{document}